\documentclass{aa}
                                    
\usepackage{graphicx}
\usepackage{txfonts}
 
\newcommand{\be}{\begin{equation}}
\newcommand{\ee}{\end{equation}}
 
\begin{document}
 
\title{Three-fluid plasmas in star formation\\
I. Magneto-hydrodynamic equations}

\author{Cecilia Pinto\inst{1}, Daniele Galli\inst{2} \and Francesca Bacciotti\inst{2}}

\titlerunning{Three-fluid MHD equations}
\authorrunning{Pinto et al.}

\offprints{C. Pinto}

\institute{Dipartimento di Astronomia e Scienza dello Spazio, 
Universit\`a di Firenze, Largo E. Fermi 5, I-50125 Firenze, Italy\\
\email{cecilia@arcetri.astro.it}
\and
INAF-Osservatorio Astrofisico di Arcetri, Largo E. Fermi 5, I-50125 Firenze, Italy\\
\email{galli@arcetri.astro.it, fran@arcetri.astro.it}
}

\date{Received ; accepted}

\abstract
{Interstellar magnetic fields influence all stages of the process of
star formation, from the collapse of molecular cloud cores to the
formation and evolution of circumstellar disks and protostellar jets.
This requires us to have a full understanding of the physical properties of
magnetized plasmas of different degrees of ionization for a wide range
of densities and temperatures.}
%
{We derive general equations governing the magneto-hydrodynamic
evolution of a three-fluid medium of arbitrary ionization, also including the possibility of charged dust grains as the main charge
carriers.  In a companion paper (Pinto \& Galli~2007), we complement
this analysis computing accurate expressions of the collisional
coupling coefficients for a variety of gas mixtures relevant for the
process of star formation.}
%
{Over spatial and temporal scales larger than the so-called large-scale
plasma limit and the collision-dominated plasma limit, and for
non-relativistic fluid speeds, the electric field, the electric current
and the ion-neutral drift have their instantaneous values determined by
the evolution of the magnetic field, which obeys an advection-diffusion
equation. The validity of the approximations made is discussed critically.}
%
{We derive the general expressions for the resistivities, the diffusion
time scales and the heating rates in a three-fluid medium and we use
them to estimate the evolution of the magnetic field in molecular
clouds and protostellar jets. Collisions between charged particles
significantly increase the value of the Ohmic resistivity during the
process of cloud collapse, affecting in particular the decoupling of
matter and magnetic field and enhancing the rate of energy dissipation.
The Hall resistivity can take larger values than previously found when
the negative charge is mostly carried by dust grains. In weakly- or
mildy-ionized protostellar jets, ambipolar diffusion is found to occur
on a time scale comparable to the dynamical time scale, limiting the
validity of steady-state and nondissipative models to study the jet's
structure.}
%
{}


\maketitle

\section{Introduction} 

The interstellar medium (ISM) is a partially-ionized gas containing
substantial fractions of neutral particles, mostly H and H$_2$, and
several charged atomic and molecular species. About 2\% of the mass of
the ISM is in the form of solid particles (dust grains) with sizes
ranging from $\sim 10$~\AA\ to $\sim 1$~$\mu$m that may carry electric
charge (Spitzer~1978, Nakano \& Umebayashi~1980, 1986).  The
transformation of a piece of ISM into a protostar involves variations
of several orders of magnitude in density, temperature, and ionization
fraction, resulting in a variety of interconnected environments, such
as cold dense clouds, cicumstellar disks, and protostellar jets.
Magnetic fields affect the motion of the charged component of the gas,
which in turn transfers the effects of the Lorentz force to the
neutrals by elastic collisions.

While several authors have focused on specific ionization regimes or
chemical compositions, we lack a rigorous formulation of the
magneto-hydrodynamic (MHD) equations in general valid for the variety
of situations occurring in the star formation process.
First, the standard expressions of the plasma conductivities are
generally derived neglecting the electron's mass with respect to the
ion and neutral's masses (e.g., Braginski~1965, Mitchner \&
Kruger~1973), and are not applicable to conditions where the carriers
of negative charge are dust grains rather than electrons. For example,
charged dust grains dominate over free electrons in clouds cores of
neutral density $\sim 10^4$--$10^5$~cm$^{-3}$ if a large number of very
small grains is present (Nishi, Nakano \& Umebayashi~1991), or during
the collapse of molecular clouds when the neutral density becomes
larger than $\sim 10^7$~cm$^{-3}$ (Nakano et al.~2002). Second, many of
the commonly adopted expressions for the conductivity of the ISM in
collapsing clouds and protostellar jets are derived neglecting
collisions between charged particles (e.g., Wardle \& Ng~1999; Desch \&
Mouschovias~2001; Nakano, Nishi \& Umebayashi~2002; Tassis \&
Mouschovias~2005). This so-called ``multifluid approach'' can be used
for an arbitrary number of charged species of arbitrary mass, but is
valid only for low-ionization degrees and cannot be generalized to
include collisions between charged particles in a self-consistent way
(see e.g., Kelley~1989). Benilov~(1996, 1997) derived the multifluid equations for a nonmagnetized plasma in detail, including accurate
expressions for the rate of momentum transfer by elastic and inelastic
collisions.

Following a different approach, in this paper we derive the general
equation for the evolution of the magnetic field in a gas of arbitrary
ionization degree made of positively- and negatively-charged particles
and neutrals. We also obtain the general expressions of the resistivity
coefficients, the associated timescales, and the heat generation rates
for a three-fluid plasma. In a companion paper (Pinto \& Galli~2007,
hereafter Paper~II), we compute accurate numerical and analytical
expressions for the collisional coupling coefficients that are required
to calculate the resistivity coefficients of the plasma.  As an
illustration, we apply the formalism derived in this paper to study the
evolution of the magnetic field in a molecular cloud and in a
protostellar jet with the help of simple analytical models. The same
formalism can be adopted to study other situations such as e.g.,
multicomponent radiatively-driven stellar winds (Krti\v{c}ka \&
Kub\'at~2001) and shock waves in plasmas of low-ionization degree
(Mullan~1971, Draine~1980, Draine \& McKee~1993; Guillet, Pineau 
des For\^ets \& Jones~2007).

In Sect.~\ref{three_fluid} we
develop a rigorous three-fluid theory for plasmas of arbitrary degree
of ionization and particle masses, critically discussing the
approximations made in deriving the governing equations; in
Sect.~\ref{flux_diffusion} we obtain the complete advection-diffusion
equation for the magnetic flux for an axially-symmetric system; in
Sect.~\ref{energetics} we obtain the equation for the rate of energy
generation due to friction forces between the fluid components; in
Sect.~\ref{molcloud} and in Sect.~\ref{jets} we apply the three-fluid
equations to typical conditions of molecular clouds and protostellar
jets, respectively, evaluating the relevant evolutionary timescales
and the heating rates; finally, in Sect.~\ref{conclusions}, we summarize
our conclusions.

\section{Three-fluid description of a partially-ionized gas}
\label{three_fluid}

We consider a system composed by three fluids, namely positively-charged particles (subscript $+$), negatively-charged particles ($-$),
and neutrals ($n$).  The generic species $s$ (with $s=+,-,n$) is
characterized by particle mass $m_s$, density $\rho_s$, mean flow velocity
${\bf u}_s$ and charge $q_s$, with $q_-=-Z_-e$ and $q_+=Z_+e$, where $e$
is the electron charge.  Charged particles may include electrons ($e$), 
ions ($i$) and positively- or negatively-charged dust grains ($g^+$,$g^-$). 
We define the quantity
\be 
\epsilon \equiv \frac{Z_+m_-}{Z_-m_+},
\label{def_eps}
\ee
the mass ratio of negatively- and positively-charged species. No
assumption is made on the magnitude of $\epsilon$: for a plasma of ions
and electrons, $\epsilon\ll 1$; for a plasma of ions and negatively-charged grains, $\epsilon \gg 1$; for a plasma of negatively- and
positively-charged grains, $\epsilon\approx 1$.

In the following, and in Paper~II, we specialize to the case where the
fluid components are characterized by a maxwellian velocity distribution
function with temperatures $T_s$, that is not necessarily the same for
all species. Necessary conditions for this assumption are that ({\em
i}\/) the characteristic evolution timescale of the system must be
much longer than the inverse of the collision frequency between
particles of a given species and ({\em ii}\/) the drift velocity
between particles of different species must be sufficiently small to
produce a frictional heating rate smaller than the thermal energy
density of each component divided by its self-relaxation time
(Draine~1986). The assumption of a maxwellian velocity distribution is
equivalent to assume that the stress tensor is diagonal and, therefore, to neglect viscosity and heat conduction in the fluid (see
Sect.~\ref{sub_def}).

The plasma is also electrically neutral, 
\be
\sum_s \frac{q_s}{m_s}\rho_s=0,
\ee
or, by Eq.~(\ref{def_eps}),
\be
\rho_-=\epsilon\rho_+.
\label{neutrality}
\ee
The assumption of charge neutrality is valid if the characteristic
length scale of the system is much larger than the Debye length of the
charged particles (see Sect.~\ref{sub_evolB} and Paper~II).

\subsection{Definitions}
\label{sub_def}

The mean velocity of each species averaged over its maxwellian velocity distribution 
function $f_s({\bf v}_s)$ is given by
\be
{\bf u}_s \equiv \langle{\bf v}\rangle_s 
=\frac{\int {\bf v}_sf_s({\bf v}_s)\;d{\bf v}_s}{\int f_s({\bf v}_s)\;d{\bf v}_s},
\ee
where ${\bf v}_s$ is the microscopic velocity of the particles, and $f_s({\bf v})$ 
is normalized as
\be
\int f_s({\bf v})\;d{\bf v}=1.
\ee
We also define the {\em mass fractions} of the three fluids,
\be
\xi_s\equiv \frac{\rho_s}{\rho},
\ee
with $\sum_s \xi_s=1$. The average density and velocity of the three-fluid system are 
\be
\rho\equiv\sum_s\rho_s,
\ee
and 
\be
{\bf U}\equiv{1\over\rho}{\sum_s \rho_s {\bf u}_s},
\label{def_U}
\ee 
respectively.

It is convenient to identify the random component of the microscopic
velocity of each species with respect to the average fluid velocity
${\bf U}$. Therefore, we define
\be
{\bf v}_s={\bf U}+{\bf w}_s,
\ee
so that
\be
{\bf u}_s=\langle{\bf U + w}\rangle_s={\bf U}+\langle{\bf w}\rangle_s,
\ee
where $\langle\;\rangle_s$ indicates the average performed on the
distribution function of the particles of species $s$.

Consistent with the assumption of a maxwellian velocity distribution,
the pressure tensor for each species
$P_s^{jk}=\rho_s \langle w_j w_k \rangle_s$ is diagonal,
\be
P_s^{jk}=P_s\delta_{jk},~~~\mbox{with}~~~P_s=\frac{1}{3}\rho_s\langle w^2 \rangle_s.
\ee
It is possible then to define the {\it kinetic temperature} $T_s$ 
\be
{3\over 2}k_B T_s\equiv {1\over 2}m_s\langle w^2\rangle_s,
\ee
where $k_B$ is the Boltzmann's constant, and the {\it thermal speed} $a_s$ 
for each species
\be
a_s^2\equiv \frac{k_BT_s}{m_s},
\ee
in terms of the {\em mean} random energy of each component.  Notice
that this definition of temperature is based on the microscopic
velocities of each species with respect to the average fluid, not with
respect to the mean velocity of each species.

With these definitions, the pressure of each component is then given by
\be
P_s=a_s^2\rho_s.
\ee
Since the random components ${\bf w}_s$ of the velocity of the different
species are defined  with respect to the {\em same} frame of reference
(the one in which the center of mass is at rest), the pressure 
contributions of each species can be consistently added together
to form the total pressure of the average fluid,
\be
P\equiv \sum_s P_s = a^2\rho,~~~\mbox{where}~~~a^2\equiv\sum_s \xi_s a_s^2.
\ee
We also define the total density and pressure of the charged species as 
\be
\rho_c\equiv\rho_+ +\rho_-, \qquad P_c\equiv P_+ +P_-.
\label{def_c}
\ee

We stress that this formulation, commonly adopted in plasma physics,
(see, e.g., Krall \& Trivelpiece~1973, Boyd \& Sanderson~1969), is
different from the approach adopted in the astrophysical literature
(e.g., Mouschovias~1991)where the random velocity of each species is
defined as a deviation with respect to the mean velocity of the {\em
same} species, ${\bf v}_s={\bf u}_s+{\bf w}_s^\prime$.  With the latter
definition, the derivation of the fluid equations for each species is
simplified by the fact that the average of the random components
$\langle {\bf w^{\prime}}\rangle_s$ are zero. However, the pressure and
the heat tensor defined in this way have a different meaning with
respect to the corresponding quantities in our derivation. Although the
partial pressures defined in the two approaches are nearly equivalent
when ${\bf u}_s \approx {\bf U} $, the definitions adopted in the two
cases, however, differ from a conceptual point of view, because in our
approach the random components of the velocity of each species are
defined in the same reference frame, and the partial pressures can be
added to form  physically meaningful global quantities.

By eq.~(\ref{neutrality}), the electric current is
\be
{\bf J}\equiv \sum_s \frac{q_s}{m_s}\rho_s {\bf u}_s=
-\frac{q_-}{m_-}\rho_-({\bf u}_+ -{\bf u}_-).
\ee
Following Schl\"uter~(1950, 1951), we also define an ``ambipolar current''
${\bf J}_d$ associated to the ion-neutral drift speed ${\bf u}_+ -{\bf u}_n$,
\be
{\bf J}_d\equiv -{q_-\over m_-}\rho_c({\bf u}_+ -{\bf u}_n).
\label{def_Jd}
\ee

\subsection{Equations of continuity and momentum}

Taking the zeroth and first order momenta of the Boltzmann equation for
each species, the equations of continuity and momentum read:
\be
\frac{\partial \rho_s}{\partial t} +
\nabla\cdot(\rho_s {\bf u}_s)= \sum_{s^\prime}S_{ss^\prime},
\label{conts1} \ee
\begin{eqnarray}
\lefteqn{\frac{\partial}{\partial t}(\rho_s {\bf u}_s) +
\nabla\cdot(\rho_s \langle{\bf v}{\bf v}\rangle_s)=
-\rho_s\nabla{\cal V}}
\nonumber \\
& & + \frac{q_s\rho_s}{m_s}
\left({\bf E}+\frac{{\bf u}_s}{c}\times{\bf B}\right) +
\sum_{s^\prime}{\bf F}_{ss^\prime}+\sum_{s^\prime}{\bf r}_{ss^\prime},
\label{moms1}
\end{eqnarray}
where ${\cal V}$ is the gravitational potential, ${\bf E}$ is the electric
field, ${\bf B}$ is the magnetic field, ${\bf F}_{ss^\prime}$ is the
friction force (per unit volume) exerted on particles of the fluid $s$
by particles of the fluid $s^\prime$ by elastic collisions, $S_{ss^\prime}$
and ${\bf r}_{ss^\prime}$ are the rate of change of density and momentum of the
species $s$ due to inelastic collisions (chemical reactions) with particles
of species $s^\prime$.

\subsection{Elastic collisions}

The calculation of the friction force ${\bf F}_{ss^\prime}$ accounting for elastic collisions
between particles with a Maxwellian velocity distribution is reviewed in
Paper~II. The resulting expression is
\be
{\bf F}_{ss^\prime}=\alpha_{ss^\prime}({\bf u}_{s^\prime}-{\bf u}_s),
\label{fcoll}
\ee
where $\alpha_{ss^\prime}$ is the so-called {\em friction
coefficient}, related to the {\it momentum transfer rate coefficient}
$\langle\sigma v\rangle_{ss^\prime}$ by
\be
\alpha_{ss^\prime}\equiv
\frac{\rho_s\rho_{s^\prime}}{m_s+m_{s^\prime}}\langle\sigma v\rangle_{ss^\prime}.
\ee
 For elastic collisions, $\alpha_{ss^\prime}=\alpha_{s^\prime s}$,
and therefore ${\bf F}_{ss^\prime}=-{\bf F}_{s^\prime s}$. In general,
$\langle \sigma v\rangle_{ss^\prime}$ is a function of the temperatures
$T_s$ and $T_{s^\prime}$ and of the relative mean velocity $|{\bf
u}_{s^\prime}-{\bf u}_s|$ of the interacting species. Thus, ${\bf
F}_{ss^\prime}$ is a nonlinear function of the relative
mean velocity.  Accurate values of $\langle\sigma v\rangle_{ss^\prime}$
for various colliding particles, and useful fitting formulae for a wide
range of temperatures and drift velocities, are derived in Paper~II.

Another frequently used quantity is the {\em collisional drag coefficient}
\be
\gamma_{ss^\prime}\equiv\frac{\langle\sigma v\rangle_{ss^\prime}}{m_s+m_{s^\prime}},
\label{drag_coe}
\ee
which is independent on the particle's density.

\subsection{Inelastic collisions}

The expressions for the rate of change of momentum due to
inelastic collisions depend on the specific type of chemical reaction
considered. For binary reactive collisions that produce one or two
particles, Benilov~(1996) has found that
\be
{\bf r}_{ss^\prime}=\alpha_{ss^\prime}^{\rm inel.}({\bf 
u}_{s^\prime}-{\bf u}_s)+S_{ss^\prime}{\bf u}_s,
\ee
where $\alpha^{\rm inel.}_{ss^\prime}$ is a complex function of the
masses and temperatures of the two reacting species, their relative
mean velocity, and the reaction cross section (or the reaction rate).
Approximated expressions for $\alpha_{ss^\prime}^{\rm inel.}$ have been obtained by Draine~(1986) for several specific reactions, and by 
Ciolek \& Mouschovias~(1993) for reactions involving dust grains.

Within a three-fluid scheme like the one considered in this paper,
the only possible reactions between two species of particles are
recombinations of singly-charged ions with electrons, resulting in the
production of neutral particles ($i + e \rightarrow n$) and ionization reactions ($n \rightarrow i + e$) . A multifluid scheme is more appropriate when explicitly considering chemical reactions and the associated change
of mass and momentum, as, for example, in the dynamics of shock waves
in the ISM (Flower, Pineau des F\^orets \& Hartquist~1985, Draine~1986)
or the collapse of magnetized molecular clouds (Ciolek \& Mouschovias~1993).

In the following, we make the hypothesis that gas is not subject to
ionization, recombination, or other chemical reactions that contribute to
the transfer of momentum from one fluid to the other. In other words,
we restrict ourselves to the study of nonreacting MHD flows.

\subsection{Equations for the mean fluid}

Neglecting the rate of change of density and momentum of each
species by chemical reactions ($S_{ss^\prime}={\bf r}_{ss^\prime}=0$),
the equations of
continuity and momentum for the mean fluid can be easily derived summing
the equations (\ref{conts1}) and (\ref{moms1}) over the species $s$,
\be
\frac{D\rho}{Dt}+\rho\nabla\cdot{\bf U}=0,
\label{cont}
\ee
\be
\rho\frac{D{\bf U }}{Dt}
= -\nabla P - \rho\nabla{\cal V} + \frac{\bf J}{c} \times {\bf B},
\label{mom}
\ee
where $D/Dt=\partial/\partial t+{\bf U}\cdot\nabla$ is the convective
derivative for the mean fluid. This set of equations must be coupled
to Maxwell's equations
\be
{\bf\nabla}\times {\bf E}=-\frac{1}{c}\frac{\partial {\bf B}}{\partial t},
\label{faraday}
\ee
\be
{\bf\nabla}\times {\bf B}=
\frac{4\pi}{c}{\bf J}+\frac{1}{c}\frac{\partial{\bf E}}{\partial t},
\label{ampere}
\ee
and
\be
\nabla\cdot{\bf B}=0.
\label{divB}
\ee 
For a self-gravitating medium, one must also add Poisson's equation,
\be
\nabla^2 {\cal V}=4\pi G\rho.
\label{poisson}
\ee

\subsection{Evolution equation for the electric current}

The mean velocities ${\bf u}_s$ of each fluid can be expressed in terms 
of ${\bf U}$, ${\bf J}$, and ${\bf J}_d$ as
\be
{\bf u}_n={\bf U}-\frac{m_-}{q_-\rho}({\bf J}-{\bf J}_d),
\label{un}
\ee
\be
{\bf u}_+={\bf U}-\frac{m_-}{q_-\rho}\left({\bf J}+
\frac{\rho_n}{\rho_c}{\bf J}_d\right),
\label{ui}
\ee
\be
{\bf u}_-={\bf U}-\frac{m_-}{q_-\rho}\left(\frac{\rho_n}{\rho_c}{\bf J}_d
-\frac{\rho_++\rho_n}{\rho_-}{\bf J}\right).
\label{ue}
\ee
The equation for the electric current ${\bf J}$ is obtained adding
together the equations of momentum of each species (Eq.~\ref{moms1})
multiplied by $q_s/m_s$, substituting the expression for the friction
force (Eq.~\ref{fcoll}), and eliminating the velocities of each species
with the help of Eqs.~(\ref{un})--(\ref{ue}). The result is:
\begin{eqnarray}
\lefteqn{{\partial {\bf J}\over\partial t}+\nabla\cdot({\bf U J}+{\bf J U})= 
-{q_-\over m_-}\nabla(P_--\epsilon P_+)}
\nonumber \\
& &+{(1+\epsilon)q_-^2\rho_-\over m_-^2}\left({\bf E}+{{\bf U}\over c}\times {\bf B}\right)
\nonumber \\
& & +{q_-\over m_-c}\left[1-(1+\epsilon){\rho_-\over\rho}\right]{\bf J}\times{\bf B}
-{(1+\epsilon)\alpha_{-+}+\alpha_{-n}\over\rho_-}{\bf J}
\nonumber \\
& & -\epsilon\frac{q_-\rho_n}{m_-c\rho}{\bf J}_d\times{\bf B}
+\frac{\alpha_{-n}-\epsilon\alpha_{+n}}{\rho_c}{\bf J}_d.
\label{evol_J}
\end{eqnarray}
Equation~(\ref{evol_J}) generalizes the evolution equation for the electric
current derived by, e.g., Rossi \& Olbert~(1970) and Greene~(1973), by
including explicitly the effects of particle collisions.

\subsection{Evolution equation for the ambipolar current}

To obtain the equation for the ambipolar current, it is convenient to
subtract the equations for the charged particles from the equation for
the neutrals in Eq.~(\ref{moms1}). Expressing the velocities
of the single species in terms of the global fluid properties with
Eqs.~(\ref{conts1}), (\ref{mom}), (\ref{un}), (\ref{ui}) and (\ref{ue}), we obtain:
\begin{eqnarray}
\lefteqn{
\frac{\partial}{\partial t}\left(\frac{\rho_n}{\rho}{\bf J}_d\right)
+\nabla\cdot\left[\frac{\rho_n}{\rho}({\bf U}{\bf J}_d+{\bf J}_d{\bf U})\right]= 
\frac{\partial}{\partial t}\left(\frac{\rho_n}{\rho}{\bf J}\right)
+\nabla\cdot\left[\frac{\rho_n}{\rho}({\bf U}{\bf J}+{\bf J}{\bf U})\right]
} \nonumber \\
& & +{q_-\over m_-}{\bf G}_n 
-{q_-\rho_n\over m_-c\rho}{\bf J}\times{\bf B}
+{\alpha_{-n}\over\rho_-}{\bf J} 
-\frac{\alpha_{-n}+\alpha_{+n}}{\rho_c}{\bf J}_d,
\label{evol_Jd}
\end{eqnarray}
where we have defined
\be
{\bf G}_n\equiv \frac{\rho_n\nabla P-\rho\nabla P_n}{\rho}.
\label{g_def}
\ee

Equation~(\ref{evol_Jd}) can be regarded  formally as an equation for the
evolution of the ambipolar current ${\bf J}_d$, analogous to
Eq.~(\ref{evol_J}) for the electric current.  The quantity ${\bf G}_n$
in Eq.~(\ref{evol_Jd}) has the dimensions of a pressure gradient and is
a source term for the ion-neutral drift that originates from density
and/or temperature differences in the two components.

\subsection{Evolution equation for the magnetic field}
\label{sub_evolB}

\begin{table*}
\caption{Characteristic length and timescales required for the
simplification of the evolution equations for the electric and
ambipolar currents, Eq.~(\ref{evol_J}) and Eq.~(\ref{evol_Jd}). Here
the neutral density $n_n$ (in cm$^{-3}$ is the density of atomic
hydrogen and $\mu$ is the mean molecular weight in units of the
hydrogen mass.  The electron and ion densities $n_e$ and $n_i$ are in
cm$^{-3}$, the magnetic field $B$ is in G, the grain radius $r_g$ is in
units of $\mu$m, the interior density of grains is assumed to be equal
to 2~g~cm$^{-3}$.}
\begin{tabular}{lllll}
\hline
composition & $T$   & $\ell_{\rm min}$ & $\tau_{\rm coll}$ & $\tau_{\rm gyr}$ \\
            & (K) & (cm)             & (s)               & (s)              \\
\hline
H$_2$, HCO$^+$, $e$   & 10     & $5.31\times 10^5 n_e^{-1/2}$     & $1.79\times 10^{10} (\mu n_n)^{-1}$     & $5.69\times 10^{-8}B^{-1}$ \\
H$_2$, H$_3^+$, $e$   & 10     & $^{\prime\prime}$                & $2.51\times 10^9 (\mu n_n)^{-1}$        & $^{\prime\prime}$ \\
H$_2$, H$^+$, $e$     & 10     & $^{\prime\prime}$                & $2.56\times 10^9 (\mu n_n)^{-1}$     & $^{\prime\prime}$ \\
H, H$^+$, $e$         & $10^4$ & $^{\prime\prime}$                & $1.01\times 10^{8} (\mu n_n)^{-1}$      & $^{\prime\prime}$ \\
H$_2$, HCO$^+$, $g^-$ & 10     & $1.23\times 10^8 n_{g^-}^{-1/2}$ & $2.83\times 10^{15} r_g (\mu n_n)^{-1}$ & $5.23\times 10^8 r_g^3B^{-1}$ \\ 
H$_2$, $g^+$, $g^-$   & 10     & $3.60\times 10^{13} r_g^{3/2}n_{g^\pm}^{-1/2}$ & $^{\prime\prime}$ & $^{\prime\prime}$ \\
\hline
\end{tabular}
\label{conditions}
\end{table*}

Equations~(\ref{evol_J}), (\ref{evol_Jd}), Faraday's law
(Eq.~\ref{faraday}) and Amp\`ere's law (Eq.~\ref{ampere}) constitute a
set of evolution equations for ${\bf J}$, ${\bf J}_d$, ${\bf B}$, and
${\bf E}$.  However, since the time variations of these four quantities
occur on different time scales, a considerable simplification and a
deeper physical insight are possible depending on the problem at hand.
This procedure of separating the timescales has long been known, but is often formulated in a
qualitative and ad hoc manner, as pointed out by Vasyli\-{u}nas~(2005).
In this section, we examine carefully the timescales associated with
each evolution equation.

Let the characteristic length and timescales for the variation of
fluid quantities be equal to $\ell$ and $\tau \approx \ell/U$,
respectively, where $U=|{\bf U}|$ is the modulus of the average
velocity defined by Eq.~(\ref{def_U}). Consider now Eq.~(\ref{evol_J})
for ${\bf J}$.  The time derivative and the convective derivative of
${\bf J}$ on the left-hand side are both of order $\sim J/\tau$, where
$J=|{\bf J}|$.  With the help of Eqs.~(\ref{faraday}) and
(\ref{ampere}), it can be easily shown that these two terms are much
smaller than the term proportional to ${\bf E}+({\bf U}/c)\times {\bf
B}$ on the right-hand side, if
\be
\ell \gg \ell_{\rm min}\equiv (1+\epsilon)^{-1/2} \lambda_-,
\label{lmin}
\ee
where
\be
\lambda_\pm=\frac{m_\pm c}{(4\pi q_\pm^2\rho_\pm)^{1/2}}=\frac{c}{\omega_\pm},
\label{def_lambdae}
\ee
is the so-called {\em inertial length} and
\be 
\omega_\pm=\frac{(4\pi q_\pm^2\rho_\pm)^{1/2}}{m_\pm}
\ee
is the {\em plasma frequency} in the charged particles. It is easily seen 
that $\ell_{\rm min}$ is equal to the inertial length of the charged
particles of smallest mass. Notice that the condition for charge neutrality,
\be
\ell \gg \lambda_{{\rm D},\pm}=\frac{a_\pm}{\omega_\pm},
\ee
where $\lambda_{{\rm D},\pm}$ is the Debye length in the charged
particles, is automatically satisfied if Eq.~(\ref{lmin}) is satisfied,
because $a_\pm \ll c$.  When the condition (\ref{lmin}) is satisfied,
the left-hand side of Eq.~(\ref{evol_J}) can be neglected, and
Eq.~(\ref{evol_J}) becomes an instantaneous equation for ${\bf E}$ in
terms of ${\bf J}$, ${\bf J}_d$, and ${\bf B}$.

Consider now Eq.~(\ref{evol_Jd}). The time derivative and the convective
derivative of ${\bf J}_d$ on the left-hand side can be neglected with
respect to the last term on the right-hand side, proportional to ${\bf
J}_d$, if
\be
\tau \gg \tau_{\rm coll}\equiv 
\frac{(1+\epsilon)\rho_+\rho_n}{(\alpha_{+n}+\alpha_{-n})\rho}
=\frac{1+\epsilon}{\rho\gamma_{\rm AD}},
\label{tcoll}
\ee
where we have defined the ambipolar diffusion drag coefficient
\be
\gamma_{\rm AD}\equiv \gamma_{+n}+\epsilon \gamma_{-n}.
\label{def_gamma_ad}
\ee 
Thus, the characteristic timescale of the system must be larger than
an effective collision timescale.  Similarly, the time and convective
derivatives of $(\rho_n/\rho){\bf J}$ are negligible with respect to
the term proportional to ${\bf J}\times {\bf B}$ on the right-hand side
if
\be
\tau \gg \tau_{\rm gyr}\equiv |\Omega_-|^{-1},
\label{tgyr}
\ee
where 
\be
\Omega_\pm=\frac{q_\pm B}{m_\pm c}
\label{def_Omegae}
\ee
is the {\em cyclotron frequency} of charged particles (negative for
negatively charged particles) and $B=|{\bf B}|$. If conditions
(\ref{tcoll}) and (\ref{tgyr}) are satisfied, the terms containing the
space and time derivatives of ${\bf J}_d$ and ${\bf J}$ can be
neglected in Eq.~(\ref{evol_Jd}).

Finally, a comparison of Faraday's law (Eq.~\ref{faraday}) and
Amp\`ere's law (Eq.~\ref{ampere}), shows that the time derivative of
${\bf E}$ (the displacement current) in Eq.~(\ref{ampere}) is of order
$(\ell/c\tau)^2\approx (U/c)^2$ with respect to the curl of ${\bf B}$.
Thus, in the following, we will neglect the displacement current in all
applications of Amp\`ere's law to ISM conditions.

Summarizing, the different levels of approximations adopted are as follows:

\begin{enumerate}

\item The time and convective derivatives of ${\bf J}$ are negligible
on length scales larger than $\ell_{\rm min}$, the inertial length of
charged particles of the smallest mass (the so-called large-scale
plasma limit), regardless of the frequency of collisions among
particles (see Vasyli\-{u}nas~2005).

\item The time and convective derivatives of ${\bf J}_d$ are negligible
on timescales larger than $\tau_{\rm coll}$, the typical timescale of
collisions between charged and neutral particles, and $\tau_{\rm gyr}$,
the inverse of the cyclotron frequency $|\Omega_-|^{-1}$ of negative
charges (the collision-dominated plasma limit, see e.g. Dungey~1958).

\item The time derivative of ${\bf E}$ (the displacement current) is
negligible if the characteristic speed of the system satisfies $(U/c)^2
\ll 1$ (nonrelativistic or weakly- relativistic regime, see e.g., Cowling~1957).

\end{enumerate}

In Table~\ref{conditions} we list the numerical values of these length
and timescales for several chemical compositions and for physical
conditions typical of molecular clouds and protostellar jets, using the
values of the collisional coefficients determined in Paper~II.

When the conditions expressed by Eqs.~(\ref{lmin}), (\ref{tcoll}) and
(\ref{tgyr}) are satisfied, Eq.~(\ref{evol_Jd}) becomes an
instantaneous expression for the ambipolar current ${\bf J}_d$,
\be
{\bf J}_d=-\frac{q_-\rho_c}{(\alpha_{+n}+\alpha_{-n})m_-} 
\left[-\frac{m_-\alpha_{-n}}{q_-\rho_-}{\bf J}
+\frac{\rho_n}{c \rho}({\bf J}\times {\bf B})-{\bf G}_n\right],
\label{Jd}
\ee  
which, substituted in Eq.~(\ref{evol_J}), neglecting the time and convective 
derivative of ${\bf J}$ neglected, gives the instantaneous expression for ${\bf E}$,
\begin{eqnarray}
\lefteqn{{\bf E}+\frac{\bf U}{c}\times{\bf B}= 
\frac{m_-}{(1+\epsilon)q_-\rho_-}\left[\nabla(P_--\epsilon P_+)
+\left(\frac{\epsilon\alpha_{+n}-\alpha_{-n}}{\alpha_{+n}+\alpha_{-n}}\right){\bf G}_n\right]}
\nonumber \\
& & +\frac{\rho_n}{(\alpha_{+n}+\alpha_{-n})\rho c}{\bf G}_n\times {\bf B}
-\left(\frac{\rho_n}{c\rho}\right)^2
\frac{({\bf J}\times {\bf B})\times{\bf B}}{\alpha_{+n}+\alpha_{-n}} \nonumber \\
& & -\left(\frac{m_-}{q_-c\rho_-\rho}\right)\left[\frac{1-\epsilon}{\epsilon}\rho_-+
\left(\frac{\alpha_{+n}-\alpha_{-n}}{\alpha_{+n}+\alpha_{-n}}\right)\rho_n\right]{\bf J}\times{\bf B} \nonumber \\
& & +\left(\frac{m_-}{q_-\rho_-}\right)^2
\left(\alpha_{-n}+\frac{\alpha_{+n}\alpha_{-n}}{\alpha_{+n}+\alpha_{-n}}\right){\bf J},
\label{gol}
\end{eqnarray}
relating the electric field ${\bf E}$ to ${\bf J}$ and ${\bf B}$ (one of the forms of the
so-called {\em generalized Ohm's equation}).

Inserting Eq.~(\ref{gol}) in Faraday's law and using Amp\`ere's
law without displacement current, we obtain the complete evolution
equation for ${\bf B}$  in the reference system of the average fluid
\begin{eqnarray}
\lefteqn{\frac{\partial{\bf B}}{\partial t}-\nabla\times
({\bf U}\times {\bf B})
=\frac{\kappa_{\rm B}}{\rho_-}\nabla(P_--\epsilon P_+)\times\nabla\rho_-
+\nabla\times(\kappa_{\rm B}^\prime {\bf G}_n)} \nonumber \\
& & 
-\nabla\times\left(\frac{\eta_{\rm D}}{P}{\bf G}_n\times{\bf B}\right)
-\nabla\times\left\{ 
\frac{\eta_{\rm AD}}{B^2}{\bf B}\times [(\nabla\times {\bf B})\times{\bf B}]
\right\}
\nonumber \\
& & -\nabla\times \left[\frac{\eta_{\rm H}}{B}(\nabla\times{\bf B})\times {\bf B}\right]
-\nabla\times (\eta_{\rm O}\nabla\times {\bf B}), 
\label{evol_B}
\end{eqnarray}
where
\be
\kappa_{\rm B}=\frac{m_+ c}{(1+\epsilon)q_+\rho_+},
\ee
\be
\kappa_{\rm B}^\prime=\kappa_{\rm B}
\left(\frac{\epsilon\alpha_{+n}-\alpha_{-n}}{\alpha_{+n}+\alpha_{-n}}\right)
\label{etan}
\ee
\be
\eta_{\rm D}=\frac{\rho_n}{\rho}\left(\frac{P}{\alpha_{+n}+\alpha_{-n}}\right)
=\frac{P}{\rho_+\rho\gamma_{\rm AD}},
\label{etad}
\ee
\be
\eta_{\rm AD}=\frac{\rho_n^2}{4\pi\rho^2}\left(\frac{B^2}{\alpha_{+n}+\alpha_{-n}}\right)=
\frac{\rho_n B^2}{4\pi\rho^2\rho_+\gamma_{\rm AD}},
\label{etaad}
\ee
\be
\eta_{\rm H}=\frac{m_+ c B}{4\pi q_+ \rho}\left[(1-\epsilon)+
\frac{\xi_n}{\xi_+}\left(\frac{\alpha_{+n}-\alpha_{-n}}{\alpha_{+n}+\alpha_{-n}}\right)
\right],
\label{etah}
\ee
\be
\eta_{\rm O}=\frac{1}{4\pi}\left(\frac{m_+ c}{q_+\rho_+}\right)^2
\left(\alpha_{-+}+\frac{\alpha_{+n}\alpha_{-n}}{\alpha_{+n}+\alpha_{-n}}\right)=
\epsilon\frac{\gamma_{\rm O}}{4\pi}\left(\frac{m_+ c}{q_+}\right)^2,
\label{etao}
\ee
and we have defined the Ohmic drag coefficient
\be
\qquad \gamma_{\rm O}\equiv \gamma_{-+}+\frac{\xi_n}{\xi_+}
\left(\frac{\gamma_{+n}\gamma_{-n}}{\gamma_{\rm AD}}\right).
\label{def_gamma_ohm}
\ee
In the limit of very high and very low ionization, the ambipolar, Hall
and Ohm resistivities reduce to the expressions given in Appendix~A.
Equation~(\ref{evol_B}) is the fundamental evolution equation for ${\bf B}$
of the problem under the given assumptions. Let us now analyze in
detail each term of Eq.~(\ref{evol_B}) and evaluate the associated
diffusion timescale.

\subsection{Ambipolar diffusion}

We consider first the fourth term on the right-hand side of
Eq.~(\ref{evol_B}), representing the evolution of magnetic field by
ion-neutral drift or ambipolar diffusion (Mestel \& Spitzer~1956).  The
associated resistivity coefficient $\eta_{\rm AD}$ is
inversely proportional to the drag coefficient $\gamma_{\rm AD}$,
defined by Eq.~(\ref{def_gamma_ad}).  Table~\ref{tab_gamma1} shows that
the contribution of electron-neutrals collisions to $\gamma_{\rm AD}$
is much smaller than the contribution of ion-neutrals collisions:  for
$\epsilon \ll 1$, $\gamma_{\rm AD}\approx \gamma_{+n}$; conversely, if
the negative charge is carried by dust grains ($\epsilon \gg 1$), their
contribution to $\gamma_{\rm AD}$ is dominant (see e.g., Nakano et
al.~2002), and $\gamma_{\rm AD} \approx \epsilon\gamma_{-n}$. The
multifluid expression for the resistivity (see Appendix~B) leads us to
underestimate the value of $\eta_{\rm AD}$ by a factor $\xi_n^2$, and
is therefore valid only for weakly- ionized gas. In a
protostellar jet, for example, where the neutral ion fraction can be as
low as $\xi_n\approx 0.5$ (see Sect.~\ref{jets}), the actual value of
$\eta_{\rm AD}$ is a factor of $\sim 4$ and higher than that obtained
with a multifluid approach.  From Eq.~(\ref{evol_B}), the timescale
of ambipolar diffusion is
\be
t_{\rm AD}\approx 
\frac{\xi_+}{\xi_n}\left(\frac{\ell}{v_{\rm A}}\right)^2\rho\gamma_{\rm AD},
\label{tad}
\ee
where $v_{\rm A}=B/(4\pi\rho)^{1/2}$ is the Alfv\`en velocity in the
average fluid.  

\begin{table}
\caption{Momentum transfer rate coefficients $\langle\sigma v\rangle_{ss^\prime}$
and collisional drag coefficients $\gamma_{ss^\prime}$ for various chemical compositions
(see Paper~II for analytical formulae and numerical results). For collisions between 
charged particles, the value of the Coulomb logarithm has been taken equal to 17.9 and 20.6
for molecular clouds and jets, respectively.}  
\begin{tabular}{llll}
\hline
species $ss^\prime$ & $\langle\sigma v\rangle_{ss^\prime}$ & $\gamma_{ss^\prime}$       & \\
                    & (cm$^3$~s$^{-1}$)                    & (cm$^3$~s$^{-1}$~g$^{-1}$) & \\
\hline
\multicolumn{4}{c}{molecular clouds ($T=10$~K)} \\
HCO$^+$, H$_2$                     & $1.73\times 10^{-9}$  &  $3.33\times 10^{13}$ & \\
H$_3^+$, H$_2$                     & $1.99\times 10^{-9}$  &  $2.38\times 10^{14}$ & \\
H$^+$, H$_2$                       & $1.17\times 10^{-9}$  &  $2.33\times 10^{14}$ & \\
H$^+$, He                          & $1.43\times 10^{-9}$  &  $1.71\times 10^{14}$ & \\
$e$, H$_2$                         & $1.97\times 10^{-9}$  & $5.88\times 10^{14}$  & \\
$e$, He                            & $1.35\times 10^{-9}$  & $2.02\times 10^{14}$  & \\
$g^\pm$ ($r_g=10$~\AA), H$_2$      & $1.77\times 10^{-9}$  & $2.11\times 10^{11}$  & \\
$g^\pm$ ($r_g=1$~$\mu$m), H$_2$    & $1.77\times 10^{-3}$  & $2.11\times 10^8$     & \\
$g^\pm$ ($r_g=10$~\AA), He      & $1.25\times 10^{-9}$  & $2.11\times 10^{11}$  & \\
$g^\pm$ ($r_g=1$~$\mu$m), He    & $1.25\times 10^{-3}$  & $2.11\times 10^8$     & \\
$e$, H$^+$                         & $2.06$                & $1.23\times 10^{24}$  & \\
\hline
\multicolumn{4}{c}{protostellar jets ($T=10^4$~K)} \\
H$^+$, H                           & $2.33\times 10^{-8}$  & $5.96\times 10^{15}$  & \\
H$^+$, He                          & $8.97\times 10^{-10}$  & $1.07\times 10^{14}$  & \\
$e$, H                             & $1.44\times 10^{-7}$  & $8.61\times 10^{16}$  & \\
$e$, H$^+$                         & $7.48\times 10^{-5}$  & $4.47\times 10^{19}$  & \\
\hline
\end{tabular}
\label{tab_gamma1}
\end{table}

\subsection{Biermann's ``battery''}
\label{subsec_biermann}

The first two terms on the right-hand side of Eq.~(\ref{evol_B})
represent the generation of magnetic fields by electric currents
produced by pressure gradients in the charged fluid, the so-called {\it
Biermann's ``battery''} (Biermann~1950, Schl\"uter \& Biermann~1950).
The first term represents a pure plasma process, whereas the second
involves collisions with neutrals. Generation of seed magnetic fields
by Biermann's battery processes are possible only if the density and
temperature gradients of the fluid components are not aligned, a
condition that may occur in stellar interiors but is unlikely in the
ISM.  It is easy to verify that the first term is larger than the
second if the negative charge is carried by electrons, whereas the
opposite is true in the case of negatively- charged grains.

\subsection{Diamagnetic current}
\label{subsec_diamagnetic}

\begin{figure}
\resizebox{\hsize}{!}{\includegraphics{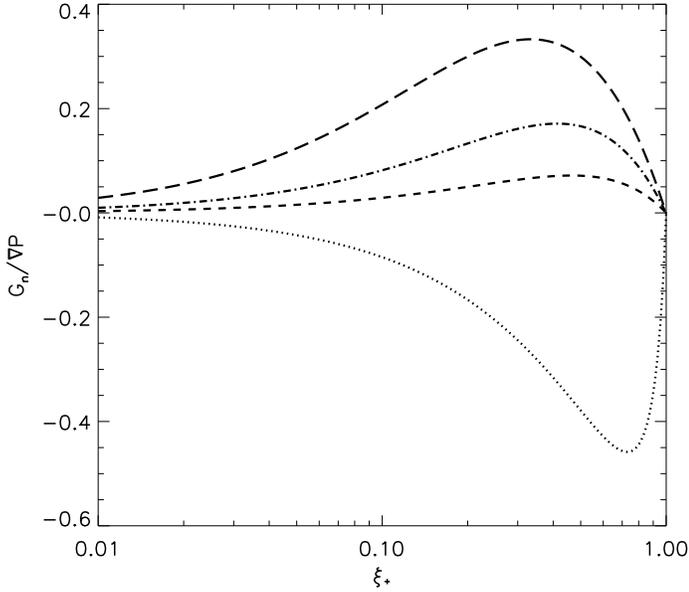}}
\caption{Function ${\bf G}_n$ normalized to the pressure gradient in the average 
fluid $\nabla P$ as function of the mass fraction of positive charges $\xi_+$. 
{\it Dotted} curve: electrons, HCO$^+$, H$_2$; 
{\it short-dashed}: electrons, H$_3^+$, H$_2$.
{\it long-dashed} curve: electrons, H$^+$, H$_2$; 
{\it dot-dashed} curve: electrons, H$^+$, H.
}
\label{gterm}
\end{figure}

The third term represents the effect of a {\it diamagnetic current},
proportional to ${\bf G}_n\times {\bf B}$, and perpendicular to the
magnetic field.  In the presence of a magnetic field and a pressure
gradient, the charges move across regions of different densities
producing a drift current perpendicular to the field and the pressure
gradient, that tends to reduce the strength of the field in the
plasma. Diamagnetic currents are generally negligible in the ISM except
perhaps near cloud's boundaries\footnote{In the geomagnetic context,
the Chapman-Ferraro current flowing at the interface between the solar
wind and the Earth's magnetopause is a diamagnetic current (see e.g.,
Parks~1991).}. There are, however, situations in which diamagnetic currents may contribute significantly to the dissipation of the magnetic field and the heating of the gas.  If the temperatures $T_s$ and the mass fractions $\xi_s$
of the fluid components are spatially constant, one finds
\be
{\bf G}_n=\xi_n(a^2-a_n^2)\nabla \rho, 
\label{g_appr}
\ee
showing that ${\bf G}_n$ vanishes both in the limit of very high
($\xi_n \rightarrow 0$) and very small ionization ($a_n \rightarrow
a$). Figure~\ref{gterm} shows the value of ${\bf G}_n$ calculated for
some gas mixtures considered in Paper~II, assuming uniform temperature
and composition.  It is clear from this figure that for intermediate
values of the ion fraction ($\xi_+\approx 0.2$--0.8, depending
on the ISM) composition, the magnitude of ${\bf G}_n$ is $\sim 40$\% of
the pressure gradient, and, therefore, is not negligible a priori.  With
${\bf G}_n$ given by Eq.~(\ref{g_appr}), the diamagnetic diffusion timescale reads
\be
t_{\rm D}\approx \frac{\xi_+}{\xi_n}
\left(\frac{\ell^2}{a^2-a_n^2}\right)\rho\gamma_{\rm AD},
\label{td}
\ee
independent on the intensity of the magnetic field.

\subsection{Hall diffusion}

The three-fluid scheme reveals that the Hall resistivity $\eta_{\rm
H}$, defined by Eq.~(\ref{etah}), is made of two terms. The first term
vanishes when the charged particles have equal mass-to-charge ratios
($\epsilon=1$), whereas the second vanishes when the particles have
equal collisional coefficients ($\alpha_{-n}=\alpha_{+n}$).  Only the
latter term appears in the multifluid approach usually adopted for a
weakly- ionized gas (see Appendix B). If $\alpha_{-n} \neq
\alpha_{+n}$, the multifluid approach is valid when the first term
inside the square parenthesis in the expression of $\eta_{\rm H}$ is
smaller than the second term, of the order of $\xi_n/\xi_+$, i.e., when
the condition
\be
\frac{\xi_+}{\xi_n} \ll \frac{1}{|1-\epsilon|}
\label{cond_hall}
\ee
is satisfied. For a plasma of neutrals, ions and electrons ($\epsilon \ll 1$), the
condition (\ref{cond_hall}) is equivalent to $\xi_+/\xi_n \ll 1$,
largely satisfied in molecular clouds and, marginally, in protostellar
jets. In this case, Eq.~(\ref{etah}) reduces to the same expression obtained 
with the multifluid approach (see Appendix~B)
\be
\eta_{\rm H}\approx \frac{\xi_n}{\xi_+}\left(\frac{m_+ c B}{4\pi q_+ \rho}\right),
\label{eta_h_e}
\ee
independent on the collisional coefficients. The associated Hall diffusion timescale is
\be
t_{\rm H}\approx
\frac{\xi_+}{\xi_n}\left(\frac{\ell}{v_{\rm A}}\right)^2\Omega_+.
\label{th}
\ee

If the negative charge is carried by ISM grains of radius $r_g$, the
multifluid expression for the Hall resistivity given by
Eq.~(\ref{eta_h_e}) is a good approximation of the correct expression
Eq.~(\ref{etah}) only when the ion fraction is extremely low,
$\xi_+/\xi_n \ll \epsilon^{-1}\approx
10^{-13}(r_g/\mbox{$\mu$m})^{-3}$. For larger mass fractions of
positive charges, typical of ISM conditions, the Hall diffusivity
becomes independent of the ion fraction,
\be
\eta_{\rm H}\approx \frac{m_- c B}{4\pi q_ \rho}.
\ee
In this case, the associated Hall diffusion time scale becomes
\be
t_{\rm H}\approx \left(\frac{\ell}{v_{\rm A}}\right)^2|\Omega_-|.
\ee
In this case, the multifluid expression for $\eta_{\rm H}$ 
given by Eq.~(\ref{eta_h_e}) {\em leads to a serious
underestimate of the actual value of the Hall resistivity}.

\subsection{Ohmic diffusion}

The Ohmic diffusion timescale is
\be
t_{\rm O}\approx 
\left(\frac{\ell}{v_{\rm A}}\right)^2\frac{\Omega_+^2}{\epsilon\rho\gamma_{\rm O}},
\ee
and is independent of the density and the intensity of the magnetic field.

\section{Advection-diffusion equation for the magnetic flux}
\label{flux_diffusion}

A more compact form of Eq.~(\ref{evol_B}) can be obtained separating the
{\it poloidal} and {\it toroidal} components of the velocity and the
magnetic field,
\be
{\bf U}\equiv {\bf U}_p + {\bf U}_\varphi, 
\qquad {\bf B}\equiv {\bf B}_p + {\bf B}_\varphi.
\ee
In the following, for simplicity, we will consider only axially-symmetric
systems ($\partial/ \partial\varphi=0$).  In a system of cylindrical
coordinated $(R,z,\varphi)$, we introduce a vector potential
\be
{\bf A}(R,z)={\Phi(R,z)\over 2\pi R} \hat{\bf e}_\varphi,
\ee
such that
\be
{\bf B}_p=\nabla\times {\bf A}=
\nabla\times \left[{\Phi\over 2\pi R} \hat{\bf e}_\varphi \right]
={1\over 2\pi R}\left(-{\partial\Phi\over\partial z}
\hat{\bf e}_R+{\partial\Phi\over\partial R}\hat{\bf e}_z\right),
\label{defB_p}
\ee
and a scalar function $\Psi$ defined as
\be
\Psi(R,z)=2\pi R B_\varphi(R,z).
\label{def_Psi}
\ee

Equation~(\ref{evol_B}) can be separated in one equation for ${\bf B}_p$,
and one equation for ${\bf B}_\varphi$.  Expressing ${\bf B}_p$ as
function of $\Phi$ according to the definition (\ref{defB_p}), the
induction equation for ${\bf B}_p$ can be ``uncurled'' and expressed in
scalar form as
\begin{eqnarray}
\lefteqn{
\frac{\partial\Phi}{\partial t}+{\bf U}_p\cdot\nabla\Phi=
\frac{\eta_{\rm D}}{P}{\bf G}_n\cdot\nabla\Phi
} 
\nonumber \\
& & 
+\frac{\eta_{\rm AD}}{(2\pi R B)^2}\left[{\cal S}(\Phi)\nabla\Phi
+\Psi\nabla\Psi\right]\cdot\nabla\Phi
\nonumber \\
& & 
+\frac{\eta_{\rm H}}{2\pi R B}{\cal J}(\Psi,\Phi)
+\eta_{\rm O}{\cal S}(\Phi),
\label{evolflux}
\end{eqnarray}
where $B=|{\bf B}|$, ${\cal S}$ and ${\cal J}$ are Stokes and Jacobi operators defined as
\be
{\cal S}(\Phi)\equiv \frac{\partial^2\Phi}{\partial R^2}
+\frac{\partial^2\Phi}{\partial z^2}-\frac{1}{R}\frac{\partial\Phi}{\partial R}.
\label{def_S}
\ee
and
\be
{\cal J}(\Psi,\Phi)\equiv \frac{\partial \Psi}{\partial R}
\frac{\partial \Phi}{\partial z}-\frac{\partial \Psi}{\partial z}
\frac{\partial \Phi}{\partial R},
\ee
respectively.

In the particular case of zero toroidal field, Eq.~(\ref{evolflux})
reduces to the simple expression
\be
\frac{\partial\Phi}{\partial t}+{\bf U}_p\cdot\nabla\Phi=
\frac{\eta_{\rm D}}{P}{\bf G}_n\cdot\nabla\Phi
+(\eta_{\rm AD}+\eta_{\rm O}){\cal S}(\Phi),
\label{evolflux_0}
\ee
In this reduced form, and neglecting the diamagnetic term,
Eq.~(\ref{evolflux_0}) is at the basis of many studies of magnetic flux
evolution by ambipolar and Ohmic diffusion in molecular clouds (see
e.g., Desch \& Mouschovias~2001).

Ambipolar diffusion produces a time variation of the magnetic flux
enclosed in a given region, at a rate depending on the ionization
fraction and the strength of the poloidal {\em and} toroidal field. Consider, for example, the realistic case of a vertical
magnetic field concentrated toward the $z$ axis, for which
$\Phi=\Phi(R)\propto R^n$ with $0<n<2$. Since ${\cal S}(\Phi)<0$ for
this configuration, ambipolar diffusion will drive the field outward of
the region under consideration, thus reducing the magnetic
flux locally.  A toroidal field will speed up (slow down) this process of field
diffusion, depending on whether the quantity $\Psi$ increases
(decreases) outward, or in other words, whether the toroidal field
decreases slower (faster) than $R^{-1}$. A toroidal field
$B_\varphi\propto R^{-1}$ has no effect on the rate of flux loss driven
by ambipolar diffusion.

The effect of the Ohmic diffusion term is to decrease (increase) the
magnetic flux enclosed in a given region if ${\cal S}(\Phi)$ is
negative (positive).  For example, for a magnetic flux
$\Phi=\Phi(R)\propto R^n$, the function ${\cal S}(\Phi)$ is positive
(negative) if $n>2$ ($n<2$), or, in other words, if the magnetic field
strength increases (decreases) outwards.  The presence of a toroidal
field has no effect on the variation of the magnetic flux by Ohmic
diffusion.

\section{Rate of energy generation}
\label{energetics}

The energy generated by the friction between streaming particles is a
significant source of heating for molecular clouds (Scalo~1977, Lizano
\& Shu~1987) and protostellar winds and jets (Ruden, Glassgold \&
Shu~1990, Safier~1993). The energy produced heats the bulk of the gas,
increasing its pressure and allowing chemical reactions to proceed at a
faster rate (Flower, Pineau des For\^ets \& Hartquist~1985, Pineau des
For\^ets et al.~1986).

In the absence of external forces, the rate of energy generation (per
unit time and unit volume) is the sum of the work done by friction
forces on the three species,
\begin{eqnarray}
\lefteqn{
W=\sum_s {\bf u}_s\cdot\sum_{s^\prime\neq s}{\bf F}_{ss^\prime}
}\nonumber \\
& & =\alpha_{-n}|{\bf u}_--{\bf u}_n|^2
+\alpha_{+n}|{\bf u}_+-{\bf u}_n|^2
+\alpha_{-+}|{\bf u}_--{\bf u}_+|^2.
\label{work}
\end{eqnarray}
Substituting in Eq.~(\ref{work}) the expressions for the drift velocities
in terms of ${\bf J}$ and ${\bf J}_d$ obtained from Eqs.~(\ref{un})--(\ref{ue}),
and eliminating ${\bf J}_d$ with the help of Eq.~(\ref{Jd}), we obtain
\be
W=\frac{1}{\rho_+\rho_n\gamma_{\rm AD}}\left|\frac{\rho_n}{4\pi\rho}[{\bf B}
\times (\nabla\times {\bf B})]
+{\bf G}_n\right|^2+\frac{\eta_{\rm O}}{4\pi}|\nabla\times {\bf B}|^2
\label{hrate1}
\ee
(see also Braginskii~1965).The first term on the right-hand side
represents the dissipation of drift motions between the plasma and the
neutrals driven by the Lorentz force or pressure gradients, whereas the
second term represents the heating of the gas due to the dissipation of
the electric current.  The ratio of the second to the first term inside
the modulus can be estimated with the help of Eq.~(\ref{g_appr}) as
\be
\frac{4\pi\rho|{\bf G}_n|}{\rho_n|{\bf B}\times (\nabla\times {\bf B})|} 
\approx \frac{a^2-a_n^2}{v_{\rm A}^2}.
\label{g_rat}
\ee
Equation~(\ref{g_rat}) shows that the contribution of pressure effects to
the heating of the gas is important only in weakly-magnetized and/or
hot plasmas with significant ionization. As shown in Fig.~\ref{gterm},
the pressure-driven drifts, represented by ${\bf G}_n$ in
Eq.~(\ref{hrate1}) are significant only at intermediate levels of
ionization.

If one sets ${\bf G}_n=0$, the heating rate is 
\begin{eqnarray}
\lefteqn{
W=\frac{1}{16\pi^3 R^2}\left[
\eta_{\rm AD}\frac{|{\cal S}(\Phi)\nabla\Phi
+\Psi\nabla\Psi|^2+|{\cal J}(\Psi,\Phi)|^2}{|\nabla\Phi|^2+|\nabla\Psi|^2} \right.
} \nonumber \\
& & \left.
+\eta_{\rm O}(|{\cal S}(\Phi)|^2+|\nabla\Psi|^2)\right].
\label{hrate2}
\end{eqnarray}
In the particular case of zero toroidal field, Eq.~(\ref{hrate2}) reduces to the simple
expression
\be
W=\frac{\eta_{\rm AD}+\eta_{\rm O}}{16\pi^3 R^2}|{\cal S}(\Phi)|^2.
\label{hrate2_0}
\ee

\section{Magnetic field diffusion in molecular clouds}
\label{molcloud}

To illustrate the formalism developed in the previous sections, we
analyze the problem of determining the diffusion of the magnetic field
in two different astrophysical environments: cold, weakly-ionized
molecular cloud cores, and hot, mildly-ionized protostellar jets.  In
this section, we analyze the case of molecular clouds, while
protostellar jets are considered in Sect.~\ref{jets}.

In dense molecular clouds with $T\approx 10$~K, $n_n\approx
10^5$--$10^6$~cm$^{-3}$, $n_i\approx 10^{-1}$--$10^{-2}$~cm$^{-3}$ and
$B\approx 10^{-5}$~G, electrons are in general the main carriers of
negative charge (Nakano et al.~2002).  In these conditions, the ion
density is determined, in a three-fluid scheme, by a balance of
cosmic-ray ionization and ion recombination, resulting in the simple
relation
\be
\rho_i\approx C\rho_n^{1/2}.
\label{ion_law}
\ee
In nondepleted clouds, the dominant ionic species are metal ions (like
Mg$^+$, Na$^+$, etc.), molecular ions (like HCO$^+$), or H$^+$ and
H$_3^+$ ions, the simple Eq.~(\ref{ion_law}) is in good agreement with
the results of detailed chemical models (Nakano et al.~2002) with
$C\approx 3.9\times 10^{-17}$~g$^{1/2}$~cm$^{-3/2}$.  In completely
depleted cores, where all metal species are frozen onto grains, the
dominant ions are H$_3^+$, H$^+$ and their deuterated analogues. In
this case, the ion density can be approximated by Eq.~(\ref{ion_law})
with $C\approx 1.5\times 10^{-17}$~g$^{1/2}$~cm$^{-3/2}$, although the
actual ion density increases with increasing density slightly less
steeply than $\rho^{1/2}$ (Walmsley et al.~2004). In
Table~\ref{tab_gamma2}, we list the values of the ambipolar diffusion
coefficients for collisions with H$_2$ appropriate for different
chemical compositions and degrees of depletion (we ignore the 10\% correction due to collisions with He, see  Paper~II).

\begin{table}
\caption{Ambipolar diffusion coefficients $\gamma_{\rm AD}$ and $\chi$
for collisions with H$_2$ in molecular clouds. The values of $\chi$ are
obtained with $C=3.9\times 10^{-17}$~g$^{1/2}$~cm$^{-3/2}$ and
$C=1.5\times 10^{-17}$~g$^{1/2}$~cm$^{-3/2}$ for non-depleted and
depleted clouds, respectively.}
\begin{tabular}{lll}
\hline
composition & $\gamma_{\rm AD}$ & $\chi$ \\
            & (cm$^3$~s$^{-1}$~g$^{-1}$) &        \\
\hline
\multicolumn{3}{c}{undepleted cloud} \\
H$_2$, HCO$^+$, $e$                     & $3.33\times 10^{13}$   & 1.0 \\
H$_2$, HCO$^+$, $g^-$ ($r_g=10$~\AA)    & $6.97\times 10^{13}$   & 2.1 \\
\hline
\multicolumn{3}{c}{depleted cloud}  \\
H$_2$, H$_3^+$, $e$                     & $2.38\times 10^{14}$   & 2.8 \\
H$_2$, H$^+$, $e$                       & $2.33\times 10^{14}$   & 2.7 \\
H$_2$, H$^+$, $g^-$ ($r_g=10$~\AA)      & $1.29\times 10^{15}$   & 15  \\
\hline
\end{tabular}
\label{tab_gamma2}
\end{table}

It can be easily verified from Table~\ref{conditions} that the
conditions for the validity of our three-fluid MHD approach, expressed
by Eqs.~(\ref{lmin}), (\ref{tcoll}) and (\ref{tgyr}) are largely
satisfied. However, a complication arises at high densities
($n_n\approx 10^{11}$--$10^{12}$~cm$^{-3}$) and relatively small scales
characterizing the so-called decoupling stage of matter and of magnetic
fields, where the electric charge is mostly carried by dust grains with
$n_{g^-} \approx n_{g^+} \approx 10^{-11}n_n$ (Desch \&
Mouschovias~2001, Nakano et al.~2002). A comparison between the typical
size of a molecular cloud during the dynamical collapse, $\ell\approx
10\,(n_{\rm n}/\mbox{cm$^{-3}$})^{-1/2}$~pc (Nakano et al.~2002), with
the value of $\ell_{\rm min}$ listed in Table~\ref{conditions} for a
plasma of negatively- and positively-charged grains with radius $r_g$
shows that the condition $\ell \gg \ell_{\rm min}$ is satisfied for
$n_{g^-}/n_n \gg 10^{-12} (r_g/\mbox{$\mu$m})^3$. This condition is
violated if grain coagulation during cloud collapse is rapid, and the
mean grain radius can reach $\mu$m-size dimensions (Flower, Pineau des
For\^ets \& Walmsley~2005).  Thus, an accurate analysis of the
decoupling stage must take into account the evolution, during the
collapse, of their mean radius to ensure that conditions (\ref{lmin}) is
satisfied at all times.  Otherwise, the evolution equation for the
field, Eq.~(\ref{evol_B}) must be replaced by the evolution equation
for the current, Eq.~(\ref{evol_J}), with ${\bf B}$ given as function
of ${\bf J}$ by Faraday's equation (Eq.~\ref{faraday}).

The most important diffusive process in molecular clouds is ambipolar diffusion,
occurring on a timescale  
\be
t_{\rm AD}\approx \frac{\xi_i}{\xi_n}
\left(\frac{\ell}{v_{\rm A}}\right)^2\gamma_{\rm AD}\rho,
\ee
(Mestel \& Spitzer~1956).  It is easy to verify that the timescales of
the Biermann's ``battery'' and diamagnetic diffusion are too long in
molecular cloud conditions.  In typical molecular cloud conditions, if
the negative charge is carried by electrons, the timescale for this
process is of the order of $\sim 10^4 (\xi_n/\xi_i)$ times the
ambipolar diffusion timescale, and therefore can be safely ignored.
Also, Biermann's battery terms identically vanish when the charge is
carried by positively- and negatively-charged grains of equal
properties. However, the Biermann's battery may lead to the generation
of seed magnetic fields at the boundary between a dense atomic or
molecular cloud and a hot diffuse external medium where gradients in
the density, temperature or chemical composition can be strong
(Lazarian~1992). For diamagnetic effects, a comparison of
Eqs.~(\ref{tad}) and (\ref{td}) shows that
\be
\frac{t_{\rm D}}{t_{\rm AD}}\approx \frac{v_{\rm A}^2}{a^2-a_n^2},
\label{tdtad}
\ee
a large quantity in weakly-ionized molecular clouds where
$(a^2-a_n^2)/a^2 \approx 2\xi_i(m_n/m_i) \ll 1$ and $v_{\rm A} \gtrsim a$.
 
In general, however, the Hall term in Eq.(\ref{evolflux}) cannot be
neglected with respect to ambipolar diffusion (Wardle~1999). If the
low-ionization condition (\ref{cond_hall}) is satisfied, we obtain from
Eqs.~(\ref{tad}) and (\ref{th})
\be
\frac{t_{\rm H}}{t_{\rm AD}}\approx \frac{\Omega_i}{\rho\gamma_{\rm AD}},
\label{thtad}
\ee
independent on the ion fraction ( Wardle \& Ng~1999). The quantity on the right-hand
side of Eq.~(\ref{thtad}) is called the ion's {\em Hall parameter} (see
Appendix~B). Inserting the values of $\gamma_{\rm AD}$ listed in
Table~\ref{tab_gamma2}, we obtain
\be
\frac{t_{\rm H}}{t_{\rm AD}}\sim 10^6 \left(\frac{B}{\mbox{$\mu$G}}\right)
\left(\frac{n_n}{\mbox{cm$^{-3}$}}\right)^{-1},
\ee
for depleted and undepleted clouds alike. Thus, for densities $n_n\sim
10^6\, (B/\mbox{$\mu$G})^{-1}$~cm$^{-3}$ or larger, the influence of Hall
diffusion competes with ambipolar diffusion in determining the
evolution of the magnetic field in a molecular cloud. The importance of 
Hall diffusion is even larger when the negative charge is carried by dust 
grains and the condition (\ref{cond_hall}) is not satisfied. In this case, 
the ratio of Hall and ambipolar diffusion timescales is
\be
\frac{t_{\rm H}}{t_{\rm AD}}=\frac{\xi_n}{\xi_i}\left(\frac{|\Omega|_-}{\rho\gamma_{\rm AD}}\right)
\approx 10^{-11}\frac{\xi_n}{\xi_i}
\left(\frac{B}{\mbox{$\mu$G}}\right)
\left(\frac{n_n}{\mbox{cm$^{-3}$}}\right)^{-1}
\left(\frac{r_g}{\mbox{$\mu$m}}\right)^{-5},
\ee
and may become of order unity or less at sufficiently high densities.
The effects of Hall diffusion have yet to be incorporated in realistic 
models of cloud collapse. The importance of the Hall diffusion for molecular clouds (and circumstellar disks) has been stressed by Wardle and coworkers
(Wardle~1999, Wardle \& Ng~1999, Pandey \& Wardle~2007). The evolution of
the cloud's magnetic field under Hall diffusion is a complex nonlinear
process, strongly coupling the poloidal and toroidal components to each
other (Goldreich \& Reisenegger~1992, Wardle~1999), which has yet to be
incorporated in realistic models of cloud collapse.

\begin{figure}
\resizebox{\hsize}{!}{\includegraphics{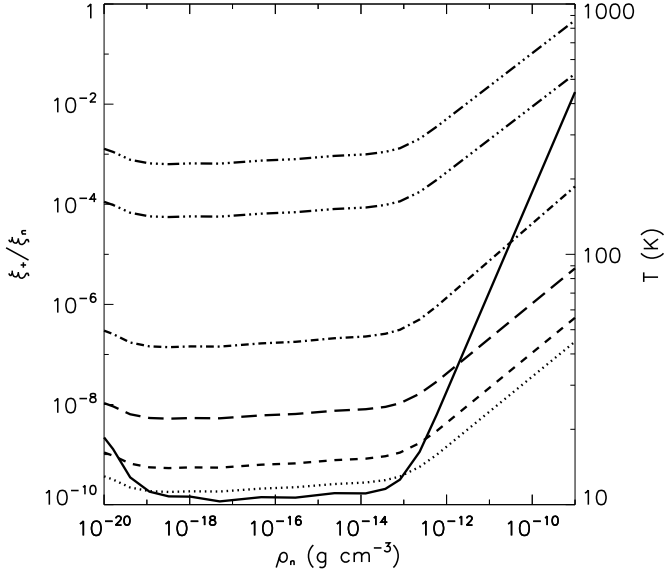}}
\caption{Importance of collisions between charged particles in molecular clouds. 
In the area above the curves, collisions between charged particles
dominate over collisions between charged particles and neutrals in
the expression of the Ohmic diffusivity $\eta_{\rm O}$.  The chemical
composition of the gas (neutrals, positive, and negative species) is:
H$_2$, H$^+$, $e$ ({\em dotted} curve);
H$_2$, H$_3^+$, $e$ ({\em short-dashed} curve);
H$_2$, HCO$^+$, $e$ ({\em long-dashed} curve);
H$_2$, HCO$^+$, $g^-$ ({\em dot-short dashed} curve);
H$_2$, $g^+$, $g^-$ ({\em dot-long dashed} curves,
with $r_g=50$~\AA\ and $r_g=100$~\AA). The temperature is shown by the
{\em solid} curve.}
\label{crit_ion}
\end{figure}

Finally, the Ohmic resistivity is often assumed to be dominated by
electron--neutral collisions (e.g., Desch \& Mouschovias~2001), because
in the multifluid treatment, where only collisions of charged
particles with the neutrals are considered, electrons have the largest
Hall parameter $\beta_{en}$ (see Appendix~B). This is hardly the case
in molecular cloud conditions, where the large value of the Coulomb
cross section compared to the polarization cross section (see Paper~II)
makes collisions between charged particles nonnegligible or even
dominant with respect to collisions with neutrals if the ionization
fraction is sufficiently high.  To illustrate this point, we show in
Fig.~\ref{crit_ion} the value of the critical ion fraction above which
the Ohmic resistivity is dominated by collisions between charged
particles, corresponding to the condition $\alpha_{-+} >
\alpha_{+n}\alpha_{-n}/(\alpha_{+n}+\alpha_{-n})$ (see
Eq.~\ref{etao}).  The critical value of the mass fraction of positive
charges $\xi_+/\xi_n$ is shown as a function of the neutral density,
assuming a temperature appropriate for a collapsing molecular cloud
core following Tohline~(1982), for various chemical compositions of the
gas, with the collisional rate coefficients given in Paper~II. For
typical conditions of dense cores, $n_n \approx
10^4$--$10^6$~cm$^{-3}$, electron--ion collisions are dominant over
electron--neutral collisions for most chemical compositions, and the
multifluid expression of $\eta_{\rm O}$ underestimates the value of
the Ohmic resistivity (see Appendix~B).  However, this limitation of
the multifluid approach has little consequence for the evolution of
molecular clouds and cores because ambipolar diffusion overwhelms
Ohmic dissipation by several orders of magnitude.  Interestingly, for
the conditions of the so-called {\it decoupling stage}, when the
magnetic field is expected to decouple from the matter in a collapsing
cloud, with $n_n\approx 10^{11}$--$10^{12}$~cm$^{-3}$, the charge is
carried mostly by dust grains with $\rho_{g^-}\approx \rho_{g^+}$
(Desch \& Mouschovias~2001, Nakano et al.~2002). Under these conditions,
depending on the grain radius, and the fraction of charged grains, the
contribution of grain-grain collisions to Ohmic dissipation may even
dominate over grain-neutral collisions, a possibility that has never
been considered in calculations of the decoupling stage of star
formation.

\subsection{A toy-model molecular cloud}
\label{cyl_cloud_model}

To estimate the timescale of magnetic flux redistribution and the
associated heating rate in the typical conditions of a molecular cloud
core, we consider a toy model for an isothermal, cylindrical gas cloud
in equilibrium under the effect of gravitational, magnetic and pressure
forces.  The hydrostatic equilibrium of the cloud is described by the
Eq.~(\ref{mom}) with $\partial/\partial t=0$ and ${\bf U}=0$, coupled
to the Poisson equation, Eq.~(\ref{poisson}). A simple solution
satisfying these equations was found by Nakamura et al.~(1993), and
reads
\be
\rho(x)=\frac{\rho_0}{(1+x^2)^2},
\ee
\be
B_R=0,~~B_z=B_0\left[\frac{1+(1-\zeta^2)x^2}{(1+x^2)^3}\right]^{1/2}, 
~~B_\varphi=B_0\frac{\zeta x}{(1+x^2)^{3/2}},
\ee
where $\rho_0$ and $B_0$ are the central values of density and magnetic
field, $\zeta$ is a parameter with value $0<\zeta< 1$ measuring the
relative importance of the toroidal and longitudinal components of the
magnetic field, and $x=R/R_0$ is the ratio of the radial coordinate $R$
and the characteristic radial scale $R_0$. The parameter $\zeta$ 
is related to $R_0$, $a$, $\rho_0$ and $B_0$ by the relation 
\be
2-\zeta^2=4\left(\frac{a}{v_{A,0}}\right)^2\left(\frac{\pi G\rho_0R_0^2}{2a^2}-1\right),
\ee
where $G$ is the gravitational constant and $v_{A,0}=B_0/(4\pi\rho_0)^{1/2}$ 
is the Alfv\`en speed on the axis of the cloud. 

\begin{figure}
\resizebox{\hsize}{!}{\includegraphics {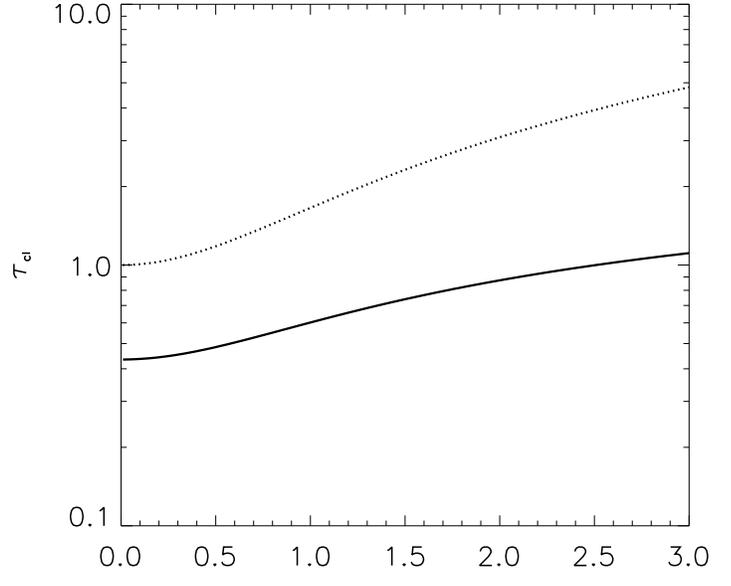}}
\caption{Timescale for ambipolar diffusion $\tau_{\rm cl}$ in the
model cloud described in Sect.~(\ref{cyl_cloud_model}) as a function of
the distance from the axis $x$ and the relative strength of the
toroidal magnetic field.  The {\em solid} curve is for $\zeta=0$, the
{\em dotted} curve for $\zeta=1$.}
\label{tcl}
\end{figure}

If the ionization fraction can be approximated by Eq.~(\ref{ion_law}), the
time scale for ambipolar diffusion in a magnetically-subcritical cloud from
Eq.~(\ref{evolflux}) is
\begin{eqnarray}
\lefteqn{ t_{\rm AD}=\frac{\chi}{t_{\rm ff,0}}\left(\frac{R_0}{2a}\right)^2
\left(\frac{\pi G\rho_0R_0^2}{2a^2}-1\right)^{-1} \tau_{\rm cl}(x,\zeta)} \nonumber \\ 
& & \approx \chi t_{\rm ff,0} \tau_{\rm cl}(x,\zeta),
\label{t_ad_cloud}
\end{eqnarray}
where
\be
\chi=\frac{\gamma_{\rm AD} C}{(8\pi G)^{1/2}}
\ee
is nondimensional parameter of order unity or larger (Shu~1983, Galli
\& Shu~1993a,b), $t_{\rm ff,0}=(2\pi G\rho_0)^{-1/2}$ is the free-fall
timescale on the axis of the cloud, and $\tau_{\rm cl}(x,\zeta)$ is
the function shown in Fig.~\ref{tcl}. In Eq.~(\ref{t_ad_cloud}), the
approximate equality is valid for strongly-magnetized clouds where
$v_{A,0} \gg a$.

For this particular model, it is easy to check that the ambipolar diffusion
term on the right-hand side of Eq.~(\ref{evolflux}) is negative for all
values of $\zeta$, and therefore the effect of ambipolar diffusion always decreases the flux contained in fluxtubes close to the cloud's
axis. The time scale of the process depends, however, on the strength of the
toroidal field.  As shown by Fig.~\ref{tcl}, the diffusion time of
the magnetic flux increases with distance from the cloud's axis and
with increasing strength of the toroidal component of the field, up to
a factor of $\sim 2$ on the cloud's axis.

The relevant collisional rate coefficients for molecular cloud
conditions computed from the expressions derived in Paper~II are listed
in Table~\ref{tab_gamma1}. Figure~\ref{tcl} shows that the evolution of
the magnetic flux (and therefore of the whole cloud) driven by
ambipolar diffusion occurs faster in the central regions of the cloud,
where $t_{\rm AD}$ is larger than the local free-fall time by a factor
$\chi$. If the innermost parts of the cloud are strongly depleted, as
often observed, (see e.g., Caselli et al.~1999), the values of the
diffusion coefficient $\gamma_{\rm AD}$ and the parameter $\chi$ listed
in Table~\ref{tab_gamma2} show that the evolution of the magnetic
fields occurs on a timescale of $\sim 3 t_{\rm ff,0}$ if the dominant
ions are H$^+$ or H$_3^+$ and the negative charge is carried by
electrons, or $\sim 15 t_{\rm ff,0}$ or larger if the negative charge
is carried by a large number of very small grains. An accurate
knowledge of the chemical composition of the central regions of a
molecular cloud core is thus necessary before conducting any evaluation of the
core's evolution timescale, if ambipolar diffusion is the driving
agent.

The ambipolar diffusion heating is 
\begin{eqnarray}
\lefteqn{W_{\rm AD}=\frac{2\rho_0}{\chi}\left(\frac{2a^2}{R_0}\right)^2
\left(\frac{\pi G\rho_0 R_0^2}{2a^2}-1\right)^2 t_{\rm ff,0}\, w_{\rm cl}(x),} \nonumber \\
& & \approx \frac{\rho_0}{2\chi t_{\rm ff,0}}\left(\frac{R_0}{t_{\rm ff,0}}\right)^2\, w_{\rm cl}(x),
\end{eqnarray}
where $w_{\rm cl}(x)=x^2/(1+x^2)^3$, and, again, the approximate
equality is valid for strongly magnetized clouds where $v_{A,0} \gg
a$.  The maximum heating rate is reached at $x=1/\sqrt{2}$, where
$w_{\rm cl}=4/27$. The Ohmic heating is always negligible
compared to ambipolar diffusion heating in typical molecular cloud
conditions.

\section{Magnetic field diffusion in protostellar jets}
\label{jets}

The diffusion of magnetic fields plays an important role in jets from
young stellar objects, caractherized by typical length $L_{\rm
jet}\approx 1$~pc or more; typical radius $R_{\rm jet}\approx 10^2$~AU; total density between $10^3$
and $10^4$~cm$^{-3}$; and temperature $T\approx 10^4$~K (Reipurth \& Bally~2001).  Recent studies have shown that the ion number
fraction $n_i/(n_i+n_n)$ can vary between 0.01 and 0.5 (Bacciotti \& Eisl\"offel~1999; Bacciotti, Eisl\"offel \& Ray 1999; Podio et al.~2006; Hartigan et al.~2007). The bulk velocity in protostellar jets is $\sim
200$~km~s$^{-1}$, so that the flow is highly supersonic (Mach number
$\sim 10$).  Little is known about the magnitude of the magnetic field,
which is difficult to determine (Mundt et al.~1990; Hartigan et al.~1994; Ray et al.~1997; Hartigan et al.~2007).
The intensity of the magnetic field in jets is sometimes
estimated to be in the range $10^2$--$10^3$~$\mu$G from equipartition
considerations.

In Table~\ref{tab_gamma1} we list the relevant collisional rate
coefficients for typical conditions of protostellar jets computed in
Paper~II.  For a jet made of H, H$^+$ and $e$ at $T=10^4$~K, we obtain
from Table~\ref{tab_gamma1} the value of the ambipolar diffusion
coefficient $\gamma_{\rm AD}= 7.01\times 10^{15}$~g$^{-1}$~cm$^3$~s,
with $e$-H collision contributing less than 1\% to the total.  For the
same composition and temperature, $\gamma_{\rm O}$ is largely dominated
by $e$-H$^+$ collisions, $\gamma_{\rm O}=4.47\times
10^{19}$~g$^{-1}$~cm$^3$~s. It can be easily verified from Table~\ref{conditions} that the conditions for the validity of our three-fluid MHD approach, expressed by Eqs.~(\ref{lmin}), (\ref{tcoll}) and (\ref{tgyr}) are largely
satisfied.

\subsection{A toy-model protostellar jet}

We apply now the equations derived in Sect.~\ref{flux_diffusion} to a
toy model of a magnetized jet, represented as a cylindrically-symmetric,
supersonic flow in equilibrium under the effect of centrifugal,
magnetic, and pressure forces.  We consider an infinite cylinder with an axis in the $z$ direction, and characterized by a radial scale $R_0$,
not necessarily coincident with the radius of the optical jet $R_{\rm
J}$. All the variables are assumed to be independent of $z$ and
$\varphi$ (cylindrical and axial symmetry), and the average flow
velocity ${\bf U}$ is everywhere parallel to the magnetic field ${\bf
B}$.  We also assume $U_R=B_R=0$ and an isothermal equation of state,
$P=a^2\rho$.

In steady state, and neglecting the gravitational potential
${\cal V}$, Eq.~(\ref{mom}) becomes
\be
\rho \frac{U_\varphi^2}{R}=\frac{{\rm d}P}{{\rm d} R}+\frac{B_\varphi^2}{4\pi R}
+\frac{1}{8\pi}\frac{{\rm d}}{{\rm d} R}(B_z^2+B_\varphi^2).
\label{eq_jet}
\ee
A particular solution of eq.~(\ref{eq_jet}) is given by 
\be
\rho(x)=\frac{\rho_0}{1+x^2},~~~{\rm with}~~~\rho_0=\frac{B_0^2}{4\pi U_{\rm jet}^2},
\ee
\be
U_z(x)=U_{\rm jet}, \qquad U_\varphi(x)= U_{\rm jet} \theta x,
\ee
\be
B_z(x)=\frac{B_0}{(1+x^2)^{1/2}}, \qquad
B_\varphi(x)= B_0\frac{\theta x}{(1+x^2)^{1/2}}, 
\ee
where $x=R/R_0$, ${\cal M}_{\rm jet}=U_{\rm jet}/a \gg 1 $ is the Mach
number of the flow along the jet, and $\theta=(1+2/{\cal M}_{\rm
jet}^2)^{1/2}$.  Since the observed Mach number in stellar jets is
${\cal M}_{\rm jet}\approx 10$, we have $\theta\approx 1$. In this
particular solution, the toroidal component of the velocity (and of the
magnetic field) increases linearly with $R$, whereas the longitudinal
velocity is independent of the distance from the axis. An upper limit
on the observed rotational velocities $U_\varphi$ in protostellar jets
is $\sim 10$\% of the axial velocity $U_{\rm jet}$ at axial distances
of $R_{\rm jet}\approx 10^2$~AU (Bacciotti et al.~2002; Woitas et
al.~2005; Coffey et al.~2007), implying that the solution above cannot
be adopted to describe an optical jet beyond $x\approx 0.1$. If
$x\approx 0.1$ corresponds to the radius of the optical jet $R_{\rm
jet}$, we must set $R_0\approx 10 R_{\rm jet}$. The density and
intensity of the longitudinal magnetic field are then almost constant
across the optically-visible jet, with the intensity of the toroidal
field increasing roughly linearly with distance from the jet's axis.

As in the case of molecular clouds, the dominant diffusive process for
the magnetic field in a protostellar jet is ambipolar diffusion.
Diamagnetic diffusion may also play a significant role at the boundary
between the jet and the ambient medium. Near the jet's axis, however,
the timescale of diamagnetic diffusion, from
Sect.~\ref{flux_diffusion}, is
\be
t_{\rm D}\approx \frac{B_0^2}{4\pi(a^2-a_n^2)\rho_0}\, t_{\rm AD}=
\frac{{\cal M}_{\rm jet}^2}{2\xi_i}\, t_{\rm AD},
\ee
more than two orders of magnitude longer than the ambipolar diffusion timescale $t_{\rm AD}$, if ${\cal M}_{\rm jet}\approx 10$. Similarly, the
timescale for Hall diffusion is
\be
t_{\rm H}\approx \frac{\Omega_+}{\rho\gamma_{\rm AD}}\, t_{\rm AD},
\ee
In the range of parameters considered, and inserting the values of $\gamma_{\rm AD}$ 
listed in Table~\ref{tab_gamma1}, we obtain
\be
\frac{t_{\rm H}}{t_{\rm AD}}\approx 10^6\left(\frac{B}{\mbox{$\mu$G}}\right)
\left(\frac{n_n}{\mbox{cm$^{-3}$}}\right)^{-1}.
\ee
The Ohmic dissipation timescale is
\be
t_{\rm O}\approx \left(\frac{\xi_n}{\xi_e}\right)\left(\frac{\gamma_{\rm AD}}{\gamma_{\rm O}}\right)
\xi_n^2\beta_{in}^2\, t_{\rm AD},
\ee
where $\beta_{in}$ is the Hall parameter of ions (see Appendix~B). For
the assumed values of jet parameters, $\beta_{in}\approx 10^5$--$10^6$,
and therefore $t_{\rm H}$ is much longer than $t_{\rm AD}$, despite the
fact that $\gamma_{\rm O}$ is about four orders of magnitude larger
than $\gamma_{\rm AD}$ (see Table~\ref{tab_gamma1}).

For $x\ll 1$ and ${\cal M}_{\rm jet}\gg 1$, 
the timescale for magnetic flux diffusion by ambipolar diffusion is
\be
t_{\rm AD}\approx \frac{1}{2}\left(\frac{\xi_i}{\xi_n}\right)
\frac{\gamma_{\rm AD}\rho_0 R_0^2}{U_{\rm jet}^2}.
\label{t_ad_jet}
\ee
Inserting order of magnitude values for the physical parameters in
Eq.~(\ref{t_ad_jet}), we obtain, on the jet's axis,
\begin{eqnarray}
\lefteqn{t_{\rm AD}\approx 6\times 10^4 
\left(\frac{\xi_i}{\xi_n}\right)
\left(\frac{\gamma_{\rm AD}}{10^{15}~\mbox{g$^{-1}$~cm$^3$~s}}\right)
\left(\frac{n}{10^{3}~\mbox{cm$^{-3}$}}\right)
\left(\frac{R_0}{10^3~\mbox{AU}}\right)^2	
} \nonumber \\
& & \left(\frac{U_{\rm jet}}{10^{2}~\mbox{km~s$^{-1}$}}\right)^{-2}~\mbox{yr},
\label{t_ad_jet_num}
\end{eqnarray}
comparable to the jet's dynamical timescale,
\be
t_{\rm dyn}=\frac{L_{\rm jet}}{U_{\rm jet}}\approx 10^4
\left(\frac{L_{\rm jet}}{\mbox{pc}}\right)
\left(\frac{U_{\rm jet}}{10^2~\mbox{km~s$^{-1}$}}\right)^{-1}~\mbox{yr}.
\ee
This important point has already been stressed by Frank et al.~(1999).
However, our expression of $t_{\rm AD}$ (Eq.~\ref{t_ad_jet_num})
differs from theirs for two reasons:  first, $t_{\rm AD}$ is
proportional to $\xi_i/\xi_n$ not to $\xi_i\xi_n$, as in the
multifluid description adopted by Frank et al.~(1999) (see
Appendix~B); second, their adopted value for the collisional rate
coefficient $\langle \sigma v\rangle_{in}$ is about one order of
magnitude smaller than the value we compute in Paper~II;
third, the curvature radius of the magnetic field $R_0$ may be larger
than the actual radius of the optical jet $R_{\rm jet}$, as shown by
our equilibrium model.

Notice that in this model $\partial\Phi/\partial t$ near the jet axis
is positive for any value of ${\cal M}_{\rm jet}$: the strong ``hoop
stress'' due to the toroidal fields compresses the plasma and the
poloidal magnetic field toward the jet's axis, where $B_z$ increases
with time over a timescale $t_{\rm AD}$.  Equation~(\ref{t_ad_jet}) shows
that the timescale of ambipolar diffusion is directly proportional to
ion fraction and the rate of mass loss in the jet, and inversely
proportional to the square of the jet's speed.  If these quantities are not strong
functions of the distance from the central star, the results of our
toy-model are robust, and indicate that the effects of ambipolar
diffusion are more important in the  outer parts of the jets, where the
dynamical timescale is longer.  In fact, Frank et al.~(1999) argued
that field diffusion already becomes effective at distances as small as
a few tens of a pc from the central source.

\begin{figure}
\resizebox{\hsize}{!}{\includegraphics{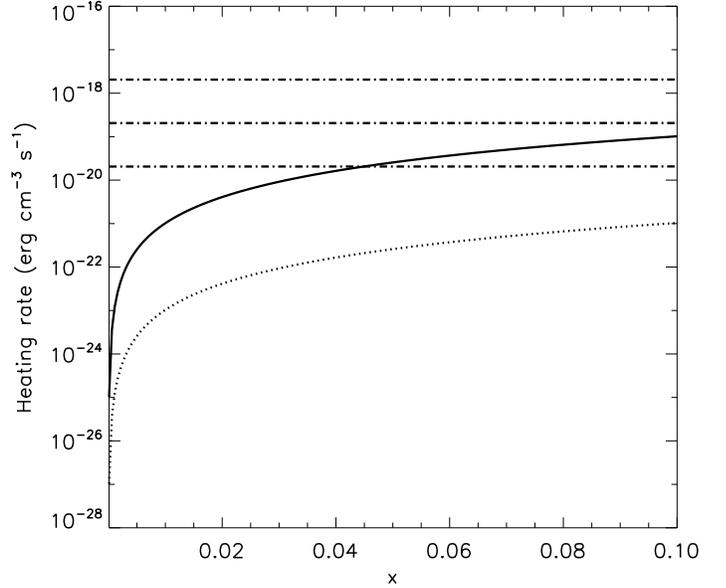}} 
\caption{Ambipolar diffusion heating rate $W_{\rm AD}$ as function of
the distance from the jet's axis $x=R/R_0$ compared with the mechanical
heating rate empirically determined by Shang et al.~(2004) for 
three values of the distance from the central star
(top to bottom: 0.01, 0.1 and 1~pc, {\em dot-dashed} lines). The {\em dotted} and
{\em solid} curves are for an ion number fraction of $n_i/(n_i+n_n)=0.5$
and 0.01, respectively.}
\label{ratejet}
\end{figure}

The ambipolar diffusion heating rate results 
\be
W_{\rm AD}=\left(\frac{\xi_n}{\xi_i}\right)
\frac{U_{\rm jet}^4}{\gamma_{\rm AD} R_0^2}\,w_{\rm jet}(x,\theta),
\label{w_ad_jet}
\ee
where $w_{\rm jet}(x)\approx x^2$ if $\theta\approx 1$.
Thus, at any fixed distance from the central source,
ambipolar diffusion heating is higher in the outer parts of the jet
rather than close to the jet's axis. The heating rate (in physical
units) for our toy-model jet as a function of the distance from the jet's
axis is shown in Fig.~\ref{ratejet}, compared with the mechanical
heating required to reproduce the observed emission of jets, estimated
following Shang et al.~(2004).  It is clear from this figure that in
weakly-ionized jets, with $n_i/(n_i+n_n)\approx 0.01$, ambipolar
diffusion heating may provide a good fraction of the required energy
input in the outer layers of the jet. These results are in qualitative
agreement with those of Garcia et al.~(2001), who considered in detail
the effects of ambipolar diffusion heating on the thermal structure of
protostellar jets. This work, however, was assuming externally a steady-state MHD model for the jet structure, not including ambipolar diffusion. The
evolution of the magnetic field internal to the jet in presence of ambipolar diffusion, however, has never been considered self-consistently in a MHD model of the jet structure and kinematics. Our work indicates instead that this point should be carefully examined in the future.

\section{Conclusions}
\label{conclusions}

We have derived a general and self-consistent form of the MHD equations
for a three-fluid system containing particles of negative and positive
charge, and neutrals. We have made no assumption about the ionization
degree, or on the particle's masses, and we have considered all
possible collisional processes between particles of the three fluids.
We have shown that the coupled equations for the electric current and
the drift current are not evolution equations above the large-scale
plasma limit and the collisionally-dominated plasma limit,
respectively, and we have derived the evolution equation for the
magnetic field valid in this regime in the reference frame of the
average fluid. The resulting expressions of the various resistivity
coefficients appearing in this equation differ from those usually
adopted in star formation studies by factors nonnegligible when the
ion fraction of the gas is significant and/or when collisions
between charged species cannot be ignored.  For axially-symmetric
systems, we have reduced the equation for the evolution of the poloidal
field to an advection-diffusion equation for the magnetic flux that
includes the effects of the toroidal field component, and evaluated the
typical timescales associated with each diffusive process. We have also
derived accurate expressions for the heating rate produced by the
collisional dissipation of drift velocities between particles of
different species driven by magnetic forces or non-uniform pressure
gradients.

We have applied our results, combined with the accurate values of
the collisional rate coefficients calculated in Paper~II, to the study
of magnetic flux dissipation in molecular clouds and protostellar
jets.  In the former case, our main conclusions are:

\begin{enumerate}

\item The timescale of ambipolar diffusion in the central regions of
magnetized clouds is of the order of the free-fall time ($\sim 1$ to 3
times depending on the degree of depletion) or about one order of
magnitude longer if the negative charge is carried by a population of
very small grains.

\item The presence of a toroidal component affects the timescale of
ambipolar diffusion by a few factors, showing that a detailed
knowledge of the magnetic field strength and morphology, in addition to
the chemical composition, is necessary for an accurate estimate of the
time scale of magnetic flux loss in molecular clouds.

\item Collisions between positively- and negatively-charged dust grains,
neglected in previous studies of cloud collapse, may significantly increase the value of the Ohmic resistivity at the high densities
where the magnetic field is expected to decouple from the gas,
enhancing the rate of the process.

\item The Hall resistivity can take larger values than previously
assumed, especially when the negative charge is mostly carried by dust
grains.

\end{enumerate}

For typical conditions of protostellar jets, characterized by higher
temperatures and ion fractions, our study shows that:

\begin{enumerate}

\item The ambipolar diffusion timescale is of the same order of
magnitude of the dynamical timescale of the jet, in agreement with
previous estimates. Thus, diffusive effects cannot be neglected in the
study of the MHD properties of protostellar jets.

\item A toy-model of a cylindrical jet shows that the hoop stresses
produced by the toroidal field are likely to force the poloidal field
to diffuse toward the jet's axis. The resulting evolution of the jet's structure has never been considered in MHD models of protostellar
jets.

\end{enumerate}

\acknowledgements
We would like to thank Prof. Claudio Chiuderi for guidance in this
work.  The research of DG and FB is partially supported by Marie Curie
Research Training networks ``Constellation" and ``Jetset", respectively.

\appendix

\section{Limiting cases}
\subsection{High ionization}

In the limit of high ionization, $\rho_n\approx 0$.  The ambipolar
diffusion resistivity, being proportional to the neutral density,
vanishes ($\eta_{\rm AD}\approx 0$), the Hall resistivity becomes
independent on collisional rate coefficients,
\be
\eta_{\rm H} \approx (1-\epsilon)\frac{m_+cB}{4\pi q_+\rho},
\label{etah_high}
\ee
and the Ohmic resistivity becomes a function of the temperature only,
\be
\eta_{\rm O}\approx \frac{1}{4\pi}\left(\frac{m_+ c}{e \rho_+}\right)^2\alpha_{-+}= 
\frac{Z_+m_- m_+c^2}{4\pi Z_-(m_- +m_+)Ze^2}\langle \sigma v\rangle_{-+}.
\label{etao_high}
\ee
In the high-ionization limit, the rate of energy generation becomes
\be
W=\frac{\eta_{\rm O}}{4\pi}|\nabla\times {\bf B}|^2.
\ee

\subsection{Low ionization}

In this limit, $\rho_+,\rho_-\approx 0$, $\rho_n\approx\rho$. The 
ambipolar diffusion and Hall resistivities can be simplified as 
\be
\eta_{\rm AD}\approx \frac{1}{4\pi}\left(\frac{B^2}{\alpha_{+n}+\alpha_{-n}}\right),
\label{etaad_low}
\ee
\be
\eta_{\rm H}\approx \frac{m_+ c B}{4\pi q_+ \rho_+}
\left[(1-\epsilon)
+\frac{\xi_n}{\xi_+}\left(\frac{\alpha_{+n}-\alpha_{-n}}{\alpha_{+n}+\alpha_{-n}}\right)\right].
\label{etah_low}
\ee

The expression for the Ohmic resistivity cannot be simplified assuming
that $\alpha_{-+}$, being proportional to the square of the ionization
fraction, is negligible with respect to $\alpha_{-n}$ and $\alpha_{+n}$
because of the large value of the Coulomb cross section compared to the
typical ion--neutral scattering cross section (see Paper~II).

In the low-ionization limit, the rate of energy generation becomes
\be
W=\frac{\eta_{\rm O}}{4\pi}|\nabla\times {\bf B}|^2+\frac{\eta_{\rm AD}}{4\pi B^2}
|{\bf B}\times (\nabla\times {\bf B})|^2.
\ee

\section{The multifluid approach}
\label{multifluid}

\begin{figure}
\resizebox{\hsize}{!}{\includegraphics{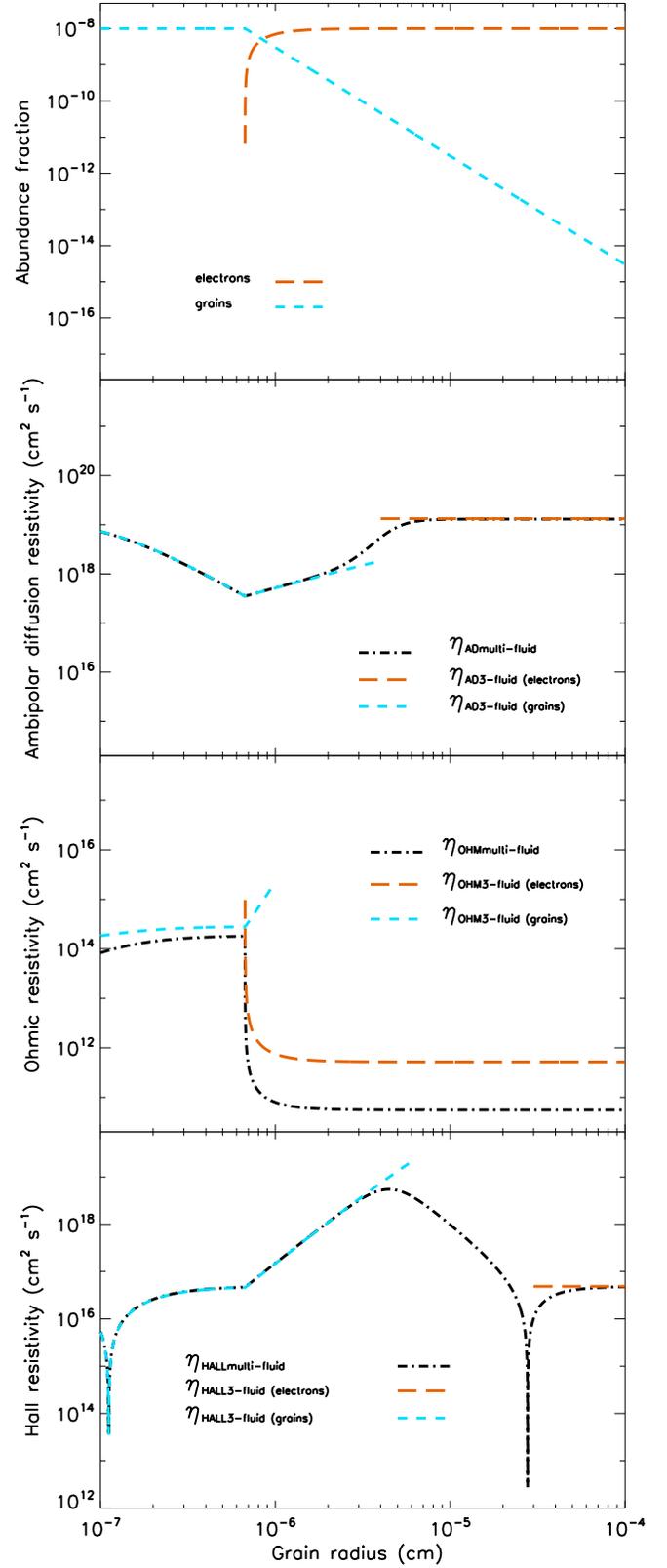}}
\caption{Values of $\eta_{\rm AD}$, $\eta_{\rm O}$ and $\eta_{\rm H}$
as function of the grain radius $r_g$ for conditions typical of a
molecular cloud core (see text).  The neutral, positive and negative
species are H$_2$, HCO$^+$, and electrons ({\em long-dashed curves}) or
grains ({\em short-dashed curves}).  The {\em dash-dotted} curves show
the same quantities computed with the multi-fluid scheme.}
\label{3eta}
\end{figure}

In this section, we compare our results with those obtained by a
multifluid scheme (arbitrary number of species, but only collisions with
neutrals included). Neglecting Biermann's battery and diamagnetic terms, 
Eq.~(\ref{gol}) can be written in the standard form
\begin{eqnarray}
\lefteqn{ {\bf E}+\frac{{\bf U}}{c}\times {\bf B}= 
\left(\frac{1}{\sigma_\parallel}-\frac{\sigma_{\rm P}}{\sigma_{\rm P}^2+\sigma_{\rm H}^2}
\right)\left[\frac{({\bf J}\times {\bf B})\times {\bf B}}{B^2}\right]
} \nonumber \\
& & +\left(\frac{\sigma_{\rm H}}{\sigma_{\rm P}^2+\sigma_{\rm H}^2}\right)\frac{{\bf J}\times {\bf B}}{B}
+\frac{\bf J}{\sigma_\parallel}
\end{eqnarray}
where $\sigma_\parallel$ is the conductivity parallel to the electric
field, $\sigma_{\rm P}$ and $\sigma_{\rm H}$ are the Pedersen and Hall
conductivities, respectively. The conductivities can be written in terms
of the {\it Hall parameters} $\beta_{sn}$ defined as
\be
\beta_{sn}=\left(\frac{q_sB}{m_sc}\right)\frac{\rho_s}{\alpha_{sn}}
=\left(\frac{q_sB}{m_sc}\right)\frac{m_s+m_n}{\rho_n\langle\sigma v\rangle_{sn}}
\ee
the ratio of the cyclotron frequency of a particle $s$ and
its characteristic frequency of collision with neutral particles (notice
that the Hall parameter is negative for negatively-charged particles).
From Eq.~(\ref{gol}), the expressions of the conductivities result
\be
\sigma_\parallel=\sigma(\beta_{+n}+|\beta_{-n}|),
\ee
\be
\sigma_{\rm P}=\sigma\left(\frac{\beta_{+n}}{1+\beta_{+n}^2}
+\frac{|\beta_{-n}|}{1+\beta_{-n}^2}\right)
\ee
\be
\sigma_{\rm H}=\sigma\left(\frac{1}{1+\beta_{+n}^2}
-\frac{1}{1+\beta_{-n}^2}\right)
\ee
where 
\be 
\sigma=\frac{q_+\rho_+ c}{m_+ B}.
\ee
These expressions are a special case, for two charged fluids, of the
general results of the multifluid approach,
\be
\sigma_\parallel=
\frac{c}{B}\sum_s \left(\frac{q_s\rho_s}{m_s}\right)\beta_{sn},
\label{mfpa}
\ee
\be
\sigma_{\rm P}=
\frac{c}{B}\sum_s \left(\frac{q_s\rho_s}{m_s}\right)\frac{\beta_{sn}}{1+\beta_{sn}^2},
\label{mfpe}
\ee
\be
\sigma_{\rm H}=
\frac{c}{B}\sum_s \left(\frac{q_s\rho_s}{m_s}\right)\frac{1}{1+\beta_{sn}^2}.
\label{mfh}
\ee

The corresponding resistivities are
\be
\eta_{\rm AD} \equiv \frac{c^2}{4\pi}\left(
\frac{\sigma_{\rm P}}{\sigma_{\rm P}^2+\sigma_{\rm H}^2}-\frac{1}{\sigma_\parallel}\right)=
\left(\frac{c^2}{4\pi \sigma}\right)\frac{\beta_{+n}|\beta_{-n}|}{\beta_{+n}+|\beta_{-n}|}.
\ee
\be
\eta_{\rm H} \equiv \frac{c^2}{4\pi}\left(\frac{\sigma_{\rm H}}{\sigma_{\rm P}^2+\sigma_{\rm H}^2}\right)
=-\left(\frac{c^2}{4\pi \sigma}\right)
\frac{\beta_{+n}-|\beta_{-n}|}{\beta_{+n}+|\beta_{-n}|},
\ee
\be
\eta_{\rm O} \equiv \frac{c^2}{4\pi \sigma_\parallel}=
\left(\frac{c^2}{4\pi\sigma}\right)\frac{1}{\beta_{+n}+|\beta_{-n}|}.
\ee
If the charged particles are electrons and ions,
$m_e\ll m_i$. Since the collisional rate coefficient are of the same order for electron-neutral
and ion-neutral collisions, in general $|\beta_{en}|\gg \beta_{in}$, and one has
\be
\eta_{\rm H}=|\beta_{en}|\eta_{\rm O}, \qquad
\eta_{\rm AD}=|\beta_{en}|\beta_{in}\eta_{\rm O}.
\ee
As an illustration, we analyze the variation of the resistivities
appearing in Eq.~(\ref{evol_B}) as function of the grain size in a
molecular cloud core.

Consider a molecular cloud core made of neutrals (H$_2$), ions
(HCO$^+$), electrons ($e$), and negatively -charged dust grains ($g^-$),
with radius $r_g$.  In particular, we assume $n_n=10^6$~cm$^{-3}$,
$n_i=10^{-2}$~cm$^{-3}$, $\rho_g/\rho=0.01$, a temperature of $T=10$~K,
and a magnetic field $B=0.1$~mG.  Each grain is assumed to be
spherical, with a radius $a$ varying between 10~\AA\ and 0.1~mm, and
mean interior density 2~g~cm$^{-3}$. All grains are assumed to be
negatively charged, and the number density of free electrons is
obtained by the neutrality condition $n_e+n_{g^-}(r_g)=n_i$.  For high
values of the grain's radius, the number density of grains is small,
and all the negative charge is carried by electrons. In the opposite
limit, the number density of small grains is high, a fraction of the
grain population is sufficient to carry all the negative charge, and
the density of free electrons is zero.

\end{document}